\documentclass[prb,twocolumn]{revtex4}

\usepackage{color}
\usepackage{array}
\usepackage{amsmath}
\usepackage{amssymb}
\usepackage{epsfig}
\usepackage{graphicx}
\usepackage{wasysym}
\usepackage{bm}
\usepackage{appendix}
\usepackage{ulem}

\begin{document}

\title{Magnetic field-dependent dynamics and field-driven metal-to-insulator transition of the half-filled Hubbard model: A DMFT+DMRG study}
\author{W. Zhu$^{1,2}$, D. N. Sheng$^{3}$, Jian-Xin Zhu$^{1,4}$}
\affiliation{$^1$Theoretical Division, T-4, Los Alamos National Laboratory, Los Alamos, New Mexico 87545, USA}
\affiliation{$^2$Center for Nonlinear Studies, Los Alamos National Laboratory, Los Alamos, New Mexico 87545, USA}
\affiliation{$^3$Department of Physics, California State University Northridge, CA 91325, USA}
\affiliation{$^4$Center for Integrated Nanotechnologies, Los Alamos National Laboratory, Los Alamos, New Mexico 87545, USA}

\begin{abstract}
We study the magnetic field driven metal-to-insulator transition in half-filled Hubbard model on the Bethe lattice,
using the dynamical mean-field theory by solving the quantum impurity problem with density-matrix renormalization group algorithm.
The method enables us to obtain a high-resolution spectral densities in the presence of a magnetic field.
It is found that the Kondo resonance at the Fermi level splits at relatively high magnetic field:
the spin-up and spin-down components move away from the
Fermi level and finally form a spin polarized band insulator.
By calculating the magnetization and spin susceptibility, we clarify that an applied magnetic field drives a
transition from a paramagnetic metallic phase to a band insulting phase.
In the weak interaction regime, the nature of the transition is continuous and captured by the Stoner's description,
while in the strong interaction regime the transition is very likely to be metamagnetic, evidenced by the hysteresis curve.
Furthermore, we determine  the phase boundary by tracking the kink in the magnetic susceptibility, and the step-like change of the entanglement entropy and the entanglement gap closing. Interestingly, the phase boundary determined from these two different ways are largely consistent with each other.
\end{abstract}

\maketitle

\tableofcontents

\section{Introduction}
The strong interplay between electron charge, spin and orbital degrees of freedom renders exotic collective behaviors,
which sensitively responds to external perturbations and thus leads to the ubiquitous quantum phase transitions in strongly-correlated systems.~\cite{Sachdev_book}
Among various ways of tuning parameters,
varying  an external magnetic field is a useful and widely applied experimental
way  for driving quantum phase transitions,
by significantly affecting the competition between the formation of the Fermi-liquid and the magnetic state  with long-range spin correlations.
The simplest model to realize such kind of magnetic field-driven phase transitions is the single-band Hubbard model \cite{hubbard63} in the presence of an exchange field.

After decades of the study, however, a consensus on the magnetic field-driven metal-to-insulator transition is yet to be reached.
Generally, there are two competing descriptions of the metal-to-insulator transition. One is the Stoner mean-field approach,~\cite{Levin83,Yoshiro_book} which
predicts a smooth magnetization change with magnetic field, while the other one is the Gutzwiller approximation predicting  a first-order metamagnetic transition
with magnetization jump around the phase transition.~\cite{Vollhardt84,Marcin14}
The following up  works largely confirm that the nature of the transition depends on the interaction strength.
That is, for the weakly correlated metal,
a smooth crossover is found between the unpolarized metal and the fully polarized band insulator, while for relatively strong interactions,
the applied magnetic field drives a first-order metamagnetic transition, forcing a jump in the magnetization curve.
Although a qualitative agreement has been achieved, the predicted critical field has non-negligible deviation between the Gutzwiller approximation and the dynamical mean-field theory (DMFT).~\cite{Laloux94,Hweson07,Parihari11}
In particular, the predicted metamagnetic transition has not been observed in experiments of liquid $^3$He,~\cite{Wiegers91}
which leaves  the nature of the field-driven metal-to-insulator transition still unsettled.
Moreover, despite that DMFT is expected to capture the dynamics and the local fluctuations better,
the obtained results are sensitive to the choice of the impurity solver.
It has been noted that, previous studies using numerical renormalization group as an impurity solver in DMFT
overestimate the transition phase boundary of the Mott transitions.~\cite{Raas04,Raas08}
Thus, it is highly desired to inspect the magnetic field-driven metal-to-insulator transition
by developing unbiased numerical techniques.

Here, we re-investigate the half-filled single-band Hubbard model with an external magnetic field.
Although a similar problem has been studied by exact diagonalization \cite{Laloux94} and numerical renormalization group \cite{Hweson07}
within the DMFT framework, the detailed behavior of the crossover from the low-field paramagnetic metal to high-field polarized insulator has not been examined systematically.
To resolve this problem,  we implement the density-matrix renormalization group (DMRG)~\cite{White92,Schollwock11}
as a quantum impurity solver into the DMFT.~\cite{Georges92a,Georges96}
The new method enables us to  provide a comprehensive picture of the crossover from low-field regime to high-field regime, with focuses on
single-particle and two-particle dynamics, as well as  entanglement measurements. It is found that the Kondo resonance at
the Fermi level splits at relatively high magnetic field, where  the spin-up and spin-down components move away from the
Fermi level and finally form a spin polarized band insulator.  In particular,
an asymmetric kink appears in the magnetic susceptibility, which signals the transition from a paramagnetic metal to a fully polarized band insulator.
With increasing the interaction, the magnetic susceptibility curve becomes steeper  approaching the transition point.
Furthermore, we also identify the phase transition by entanglement measurements.
The step-like change in entanglement entropy and entanglement gap closing faithfully represents
a quantum phase transition, with the critical field consistent with that obtained from the magnetic susceptibility.


This paper is organized as follows: In Sec. II, we start by briefly introducing the model and reviewing the general framework
of the DMFT  and the DMRG impurity solver. We close this section with a discussion on the mapping of
the usual impurity Anderson model to a two-component spinless model.
In Sec. III, we describe the magnetic field dependent dynamics of the system.
We study the evolution  of spectral densities, magnetization curves and magnetic susceptibilities with the change of  the magnetic field.
We also introduce the entanglement measurements to detect the metal-to-insulator transition.
The conclusion is given in Sec. IV.

\section{Model and Methodology}

\subsection{Hubbard model in a magnetic field}\label{sec:model}
We study the two-dimensional Hubbard model in the presence of a magnetic field:
\begin{eqnarray}
  \label{eq:ham_hubb}
  \mathcal{H} =&&
  U \sum_i \left(n_{i,\uparrow  }-\frac{1}{2}\right)
           \left(n_{i,\downarrow}-\frac{1}{2}\right) \nonumber\\
  &&- t \sum_{\langle i,j\rangle, \sigma}
    c^\dagger_{i,\sigma} c^{\phantom\dagger}_{j,\sigma} +h \sum_i (n_{i,\uparrow} - n_{i,\downarrow})
\end{eqnarray}
where $c^\dagger_{i,\sigma}$ creates a spin-$\sigma$ electron at site $i$. $t$ is the hopping matrix element between two nearest neighbor sites and $U$ is the on-site interaction. Zeeman splittng strength $h$ is determined by external magnetic field ($H$) by $h=g\mu_BH/2$ ($\mu_B$ is Bohr magneton and $g$ is Lande factor).
In the present work, we have made several  assumptions to simplify the problem: First, we consider the single orbital model. Second,
we just consider the external magnetic field leading to Zeeman effect and neglect the orbital effect from magnetic field.
Third, throughout the work, we focus on the half-filling case (or Fermi level (chemical potential) is set to be $\mu=0$.) and $t=1$.

In this paper, we study the Hubbard model, Eq.~(\ref{eq:ham_hubb}), defined  on the Bethe lattice, and solve it within DMFT.
Working on the Bethe lattice significantly simplifies the self-consistent calculations within DMFT,
as shown in Sec. \ref{sec:dmft}. To be specific, we only use the property of the Bethe lattice, whose density of states
in the absence of interactions takes the form of
\begin{equation}
  \label{eq:bethe_dos}
  \rho^0(\omega) =\frac{2}{\pi D^{2}}\sqrt{D^{2}-\omega^{2}},
\end{equation}
where $2D=4$ stands for the corresponding band width (see Sec. \ref{sec:dmft_bethe} for details).

\subsection{Dynamical mean-field theory} \label{sec:dmft}
The DMFT~\cite{Georges92a,Georges96} is a non-perturbative treatment of the electronic structure in strongly-correlated systems,
which bridges the gap between the  fully itinerant  limit and the fully localized limit of electronic states.
The key idea of DMFT is the mapping of a many-body lattice problem to an Anderson impurity model, which is solved self-consistently.
The Anderson impurity model describes the hybridization of interacting electrons located on one or several sites (the impurity sites) with a ``bath'' of conduction electrons.~\cite{Hewson93}
While the mapping  itself is exact, the approximation made in ordinary DMFT schemes is to assume the lattice self-energy to be momentum-independent (neglecting all non-local correlations),
which only becomes exact in the limit of lattices with an infinite coordination number (for example, the Bethe lattice).

Generally, the DMFT self-consistent condition is the lattice Green's function ($G_\sigma(\omega)$) coincides with impurity Green's function
($g_\sigma(\omega)$) from the Anderson impurity model by (See Appendix. \ref{sec:dmft_bethe} for details)
\begin{equation}\label{eq:self-con}
 G_\sigma(\omega)\approx g_\sigma(\omega) \;.
\end{equation}
This equation is used to set up a DMFT iteration cycle to
find a self-consistent solution. To be specific, on the Bethe lattice,
the self-consistent condition can be represented by (see Appendix \ref{sec:dmft_bethe})
\begin{equation}
  \Gamma_\sigma(\omega) = \frac{D^{2}}{4}G_\sigma(\omega).
\end{equation}

\subsection{Dynamic DMRG as impurity solver}\label{sec:dmrg}
Solving the Anderson impurity model amounts to  computing observables such as the interacting Green's function
and related spectral function for a given hybridization function.
There exists a number of ways to solve the Anderson impurity model,~\cite{Georges96,Hewson93}
including exact diagonalization,~\cite{Cakarel94,Si94,Hallberg95} the numerical renormalization group,~\cite{Wilson75,Bulla08}
iterative perturbation theory,~\cite{Georges92a,Hewson16} the Hirsch-Fye~\cite{JEHirsch:1986,XYZhang93,Rozenberg97,Jarrel92,Georges92b}
and continuous-time~\cite{PWerner:2006,EGull:2011}  quantum Monte Carlo methods. Although the above impurity solvers have been proposed and developed for decades,
they have strength and weakness.
For instance, the numerical renormalization group, being
designed for impurity problems, is unable to resolve a good
resolution of spectral density at high energy regime,
due to the limitation of logarithmic discretization of the bath density of states.
Moreover, related generalization of numerical renormalization group
to the multi-orbital or multi-band lattice model is still unfeasible.
Quantum Monte Carlo method
can efficiently deal with  multi-band models, but it lacks high resolution of the
spectral function when formulated in imaginary time, due to the ill-conditioned
analytic continuation from imaginary to real frequencies.
Exact diagonalization naturally works with real frequencies, but it
is severely limited by its accessible system sizes. This
again reduces the spectral resolution considerably.

On the other hand, over more than twenty years of the development, DMRG~\cite{White92,Schollwock11} has become
a mature numerical technique dealing with generalized Hamiltonian,
which is widely accepted as the most successful method for one-dimensional interacting systems.
Since the impurity problem in DMFT can be transformed to be actually one-dimensional,
it is natural to explore the combination of DMFT and  DMRG, and
apply DMFT+DMRG method to problems in dimension higher than one.
Along this direction,  several promising schemes have been
proposed in the past years, for example, the dynamical DMRG algorithm~\cite{White99,Jecke02,Nishimoto03,Gebhard03,Carcia04,Nishimoto04} and
the extended Chebyshev matrix-product impurity solver.~\cite{Holzner11,Dargel12,Wolf14,Ganahl15}
General advantages are that the DMRG-based impurity solver works at zero temperature and real frequency domain,
and the spectral function can be obtained  with high precision for all frequency
and calculated with uniform resolution.

In this paper, we choose the scheme proposed by Refs.~\onlinecite{White99,Jecke02}.
To be specific, the ordinary DMRG is just for targeting the ground state, saying $|0\rangle$.
To reach the dynamics of the system, generally speaking, one should know the
knowledge about excited states or excitation information.
As proposed in Ref.~\onlinecite{White99}, if only the spectral function or dynamical correlation function is concerned, one  does not need to target full excitation energy spectrum.
Instead, the spectral function can be calculated directly using a \textit{correction vector} state:
\begin{equation}
 |x^{\pm}_d(\omega+i\eta) \rangle= 	\frac{1}{\omega\pm(E_0-\hat H) +i\eta} \hat d^{\pm} |0\rangle .
\end{equation}
With the help of the correction vector, the Green's function can be calculated directly:
\begin{equation}
	G_d(\omega+i\eta) = \langle0| \hat d  | x^{+}_{d}(\omega+i\eta) \rangle + \langle 0| d^{\dagger}| x^{-}_d(\omega+i\eta) \rangle 
\end{equation}
where $\eta$ is a nonzero positive value for smearing energy (in this paper we focus on the retarded Green's function).
Taking these states ($|0\rangle$,$\hat d^\dagger|0\rangle$ and $|x^{\pm}_d(\omega) \rangle$)
as target states and optimizing the DMRG basis to represent them allow for
a very precise calculation of the Green's function for
a given frequency $\omega$ and the broadening factor $\eta$.
This is usually called  dynamical DMRG algorithm.~\cite{White99,Jecke02}

In Appendix \ref{sec:SIAM_test1} and \ref{sec:SIAM_test2}, we have performed extensive tests of the above DMRG scheme and DMFT+DMRG scheme.
Through these benchmarks, we conclude the current scheme can reach reliable dynamical properties efficiently.

\subsection{Two-component mapping of single impurity Anderson model}
\label{sec:mapping}
\begin{figure}[ht]
  \includegraphics[width=0.95\columnwidth]{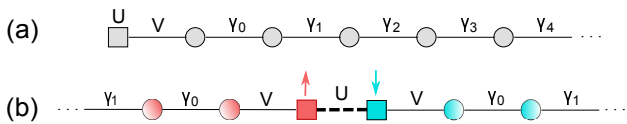}
  \caption{(a) Single impurity Anderson model with the bath as half-infinite chain.
           (b) Equivalent model with two impurities coupled with two half-infinite bath chain.
           Square and circle dot respectively represent the impurity site and bath site.}\label{fig:geom}
\end{figure}

Within the DMFT scheme, the key procedure is to map the original Hamiltonian, Eq. (\ref{eq:ham_hubb}), to a one-dimensional Anderson impurity
model.~\cite{Georges92a,Georges92b,Bulla08,Georges96}
The linear chain version of the single impurity Anderson model, where the bath of conduction
electrons is described by a hybridization function in continued fraction representation (see Appendix \ref{sec:dmft_bethe}),
can be illustrated in Fig. \ref{fig:geom} (a).
Note that the interaction term only appears on the impurity site (square dot in Fig. \ref{fig:geom} (a)).
Thus it is possible to mapping the single impurity Anderson model from a spinful model to a two-component
spinless model ($\hat c_{i,\uparrow}\rightarrow \hat a_{i}, \hat c_{i,\downarrow}\rightarrow \hat b_{i}$) as:
\begin{eqnarray}\label{eq:twospinless}
  H &=& U(n_{d,\uparrow}-1/2) (n_{d,\downarrow}-1/2) \nonumber\\
 && + V (d^{\dagger}_{\uparrow} a_{1} +h.c.) +
  \sum_{i}\varepsilon_i a^{\dagger}_{i} a_{i} + \sum_{i} \gamma_i(a^{\dagger}_{i} a_{i+1}+h.c.) \nonumber\\
 &&+ V (d^{\dagger}_{\downarrow} b_{1}+h.c.) +
  \sum_{i}\varepsilon_i b^{\dagger}_{i} b_{i} + \sum_{i} \gamma_i(b^{\dagger}_{i} b_{i+1}+h.c.), \nonumber \\
\end{eqnarray}
where $\hat d^{\dagger}_{\sigma}$ creates a  spin-$\sigma$   electron at impurity site and $\hat a^{\dagger}_{i} (\hat b^{\dagger}_{i})$
creates spinless electron in the bath \cite{Bulla08} (see Appendix \ref{sec:discre}).
After this mapping, it is straight forward to see two fermion bath are actually decoupled.
In numerical calculation, we use the geometry shown in  Fig. \ref{fig:geom} (b), where
two impurity sites interact through density-density interaction and each is coupled with one semi-infinite fermion bath respectively.   
In literature, the geometry of Fig. \ref{fig:geom} (a) is widely used since this is the specific setting for numerical renormalization group calculations \cite{Bulla08}.
For the DMRG calculation, the alternative geometry of Fig. \ref{fig:geom} (b) is a more natural choose \cite{note1}.

The mapping from the spinful Hubbard model
onto a two-component spinless model has several advantages, due to the essential realization of DMRG algorithm.
First, this two-component mapping improves the accuracy of DMRG calculations.
The DMRG is a real space variational scheme where building blocks of the
whole system are enlarged by one lattice site and then are updated at each iterative step.
After updating, the enlarged building blocks need to be projected onto a
reduced (or ``importance'') basis set, therefore the projection (truncation) error is relatively smaller
if the Hilbert space of enlarged building blocks are kept small.
In the current case,
the two-component mapping makes the original Hilbert space be the direct product of two local Hilbert space (spin-up and spin-down),
so that the fermionic Hilbert space at each site is reduced from four to two.
Thus, adding a single spinless site instead of a spinful site in each DMRG step leads to a much smaller truncation error or higher resolution.
Second, this two-component mapping saves the computational resources in DMRG calculations.
Generally, when we work on the problem with smaller local Hilbert space, the numerical cost in DMRG is exponentially reduced.
But this is not transparent here because the total Hilbert space is not changed (as we split each spinful site into two spinless sites).
In Appenidx \ref{sec:time}, we compare the computational performance of spinful Hubbard model and two-component spinless model.
It is found that the performance of the two-component spinless model is faster by an approximate factor of two.
To sum up, we conclude that two-component mapping has a great advantage for solving Anderson impurity model based on DMRG.



\section{Results and Discussion}
\subsection{Spectral density}
First of all, we examine the effect of magnetic field on spectral density function defined as
\begin{eqnarray}
   \rho(\omega)=-\frac{1}{\pi}\mathfrak{Im} G^R(\omega)=-\frac{1}{\pi} \lim_{\eta \rightarrow 0^{+}} \mathfrak{Im} G_{d}(\omega+i\eta).
\end{eqnarray}
The impurity Green function $G_{d}(\omega+i\eta)$ is obtained from DMFT+DMRG scheme:
\begin{eqnarray}
  G_d(\omega+i\eta)= \langle 0 | \hat d_{\sigma} \frac{1}{\omega+i\eta +E_0 -\hat H}   \hat d^{\dagger}_\sigma |0 \rangle + \nonumber\\
  \langle 0 | \hat d^{\dagger}_{\sigma} \frac{1}{\omega+i\eta -E_0 +\hat H}   \hat d_\sigma |0 \rangle,
\end{eqnarray}
where $E_0$ stands for the ground state energy and $\eta$ is the small broadening parameter (see Appendix  \ref{sec:decon} for details).
Figure~\ref{fig:spec} shows the spectral density for the majority (minority) spin-down (spin-up) electrons for various magnetic field $h$ by setting interaction strength $U=D$ and $U=1.5D$.
We can see that, by increasing magnetic field $h$, the more and more spectral weight of spin-up (spin-down) component is shifted to higher (lower) energy regime.
When magnetic field reaches a threshold $h\geq h_c$, the spectral density is completely spin polarized (chemical potential setting as half-filling), indicting the system becomes a polarized band insulator.

\begin{figure}[b]
  \begin{center}
  \includegraphics[width=0.49\columnwidth,angle=0]{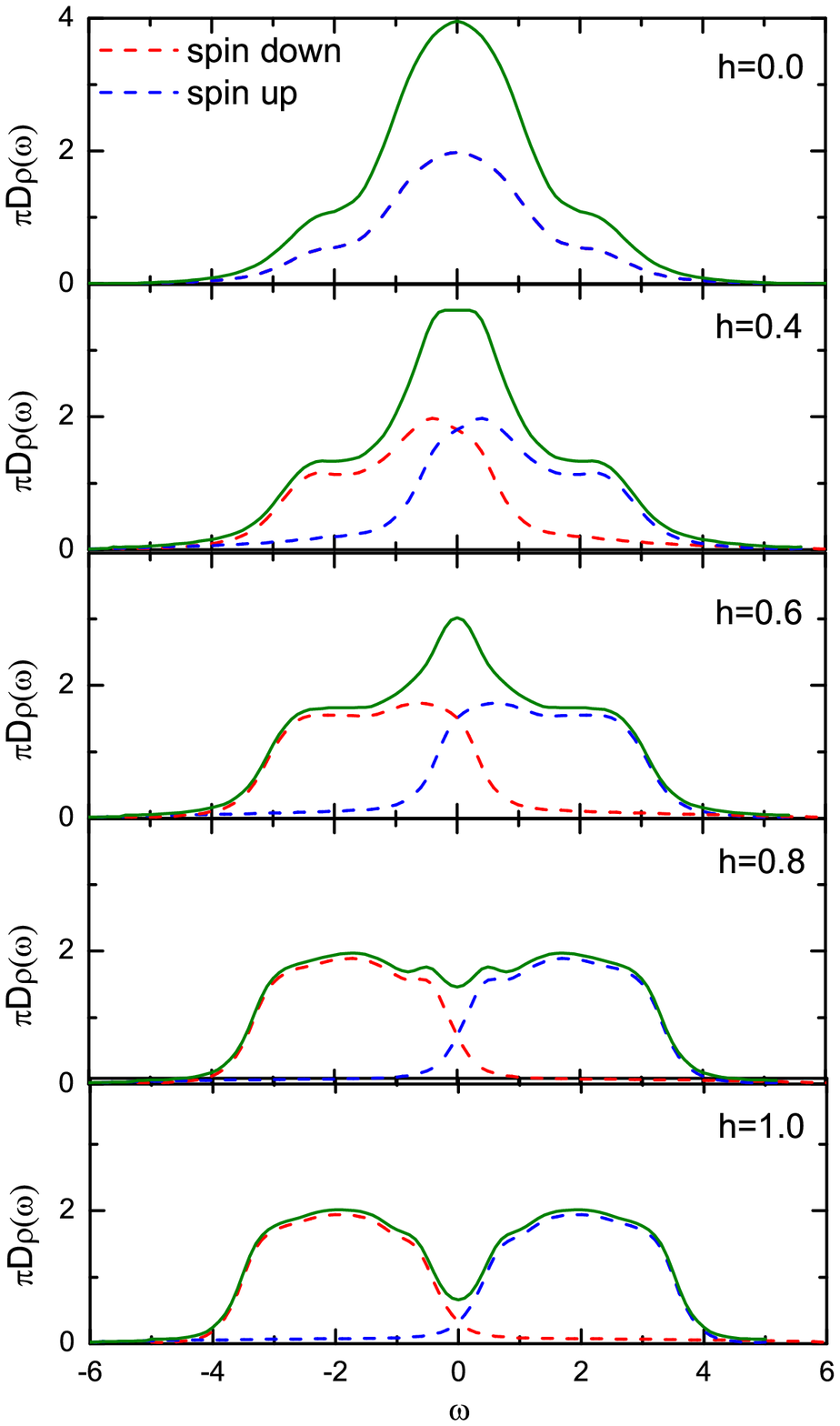}
  \includegraphics[width=0.49\columnwidth,angle=0]{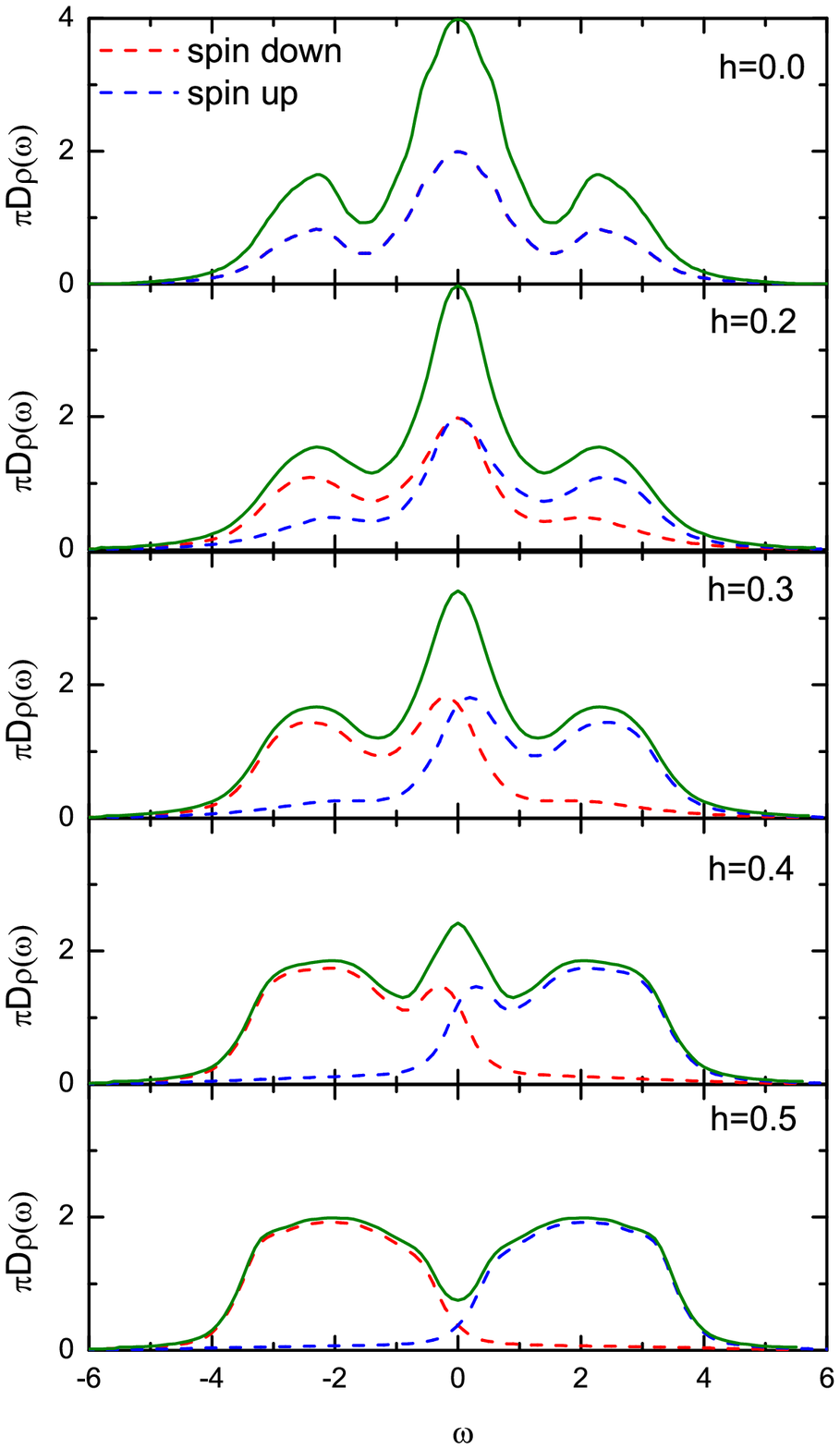}
    \caption{Spin-dependent spectral density of Bethe lattice for various values of magnetic field strength $h$: spin-down $\rho_{\downarrow}(\omega)$ (red dashed line), spin-up $\rho_{\uparrow}(\omega)$ (blue dashed line) and total density $\rho(\omega)=\rho_{\downarrow}(\omega)+\rho_{\uparrow}(\omega)$ (green solid line).
    The left column corresponds to the interaction strength $U=D=2.0$ and the right column to  $U=1.5D=3.0$.
     As to the DMFT impurity solver, we use a chain enclosing $L=80$ fermionic sites, which is solved by DMRG algorithm by keeping $128$ states.
     The DMRG projection error in each variational step is less than $10^{-8}$.
     DMFT iterations is stopped by the condition $\Delta \rho(\omega)=\rho^{i+1}(\omega)-\rho^{i}(\omega)<0.01$ for every $\omega$ ($\Delta \rho(\omega)$ is the spectral density difference between two consecutive DMFT iterations). Before the use of deconvolution scheme, the smearing energy in calculating the Green's function is set by $\eta=0.2D$.}\label{fig:spec}
  \end{center}
\end{figure}

Next we discuss the behavior of Kondo resonance peak.
It is found that the Kondo resonance peak survives in weak magnetic field. Importantly, we observe that the Kondo resonance peak of spin-up (or spin-down) is not pinned at zero frequency,
since the electron-hole symmetry is broken due to the magnetic field. The Kondo peak in the spin-up (spin-down) spectral density shifts towards low (high) frequency regime, with increasing magnetic field.
Interestingly, the total spectral density (green solid curve in Fig. \ref{fig:spec})
forms a resonance peak centered at zero frequency before entering the fully polarized regime (see below).
In high magnetic field, the Kondo resonance peak disappears, instead we observe a dip at Fermi level.

\subsection{Magnetization}
\begin{figure*}[!htb]
  \begin{center}
  \includegraphics[width=0.3\textwidth,angle=0]{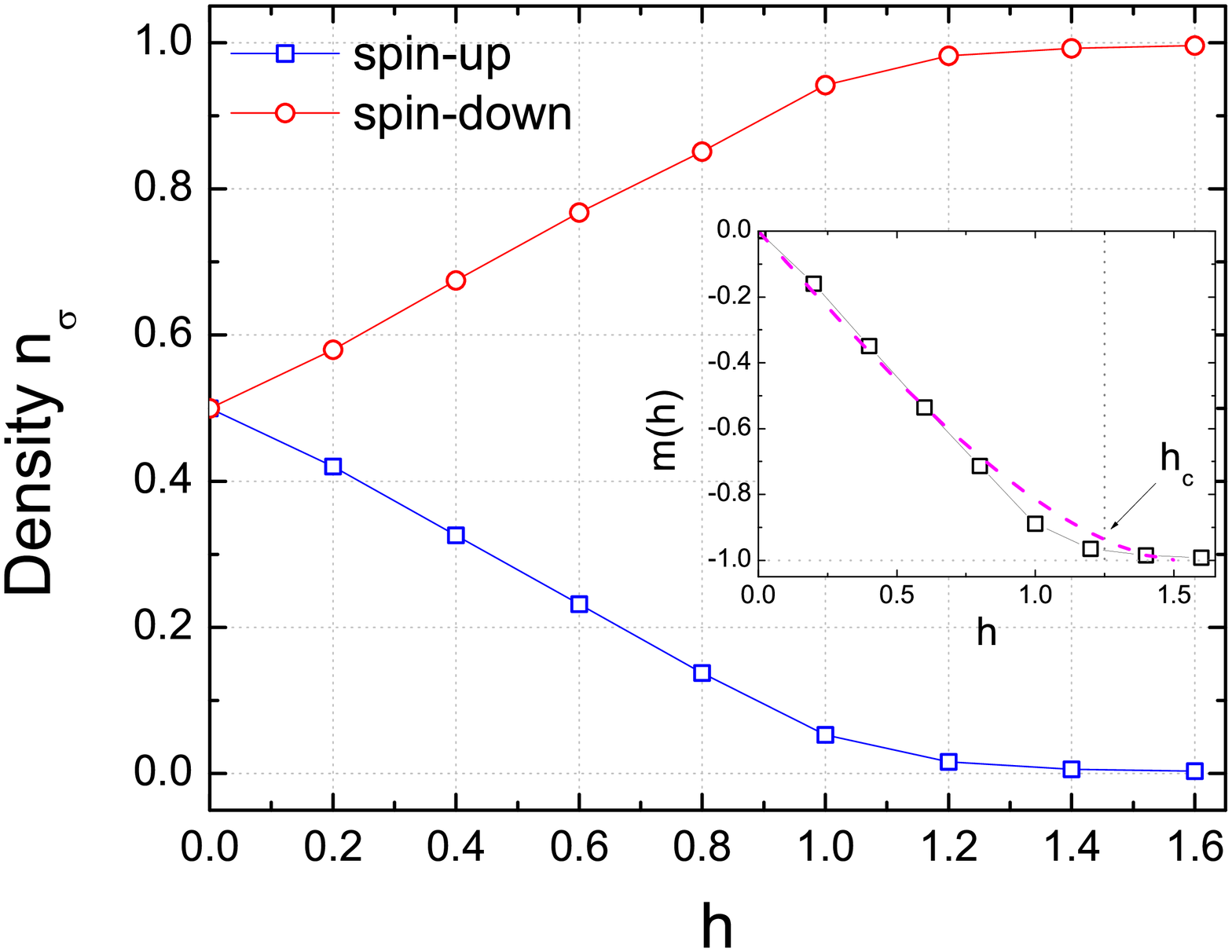}
  \includegraphics[width=0.3\textwidth,angle=0]{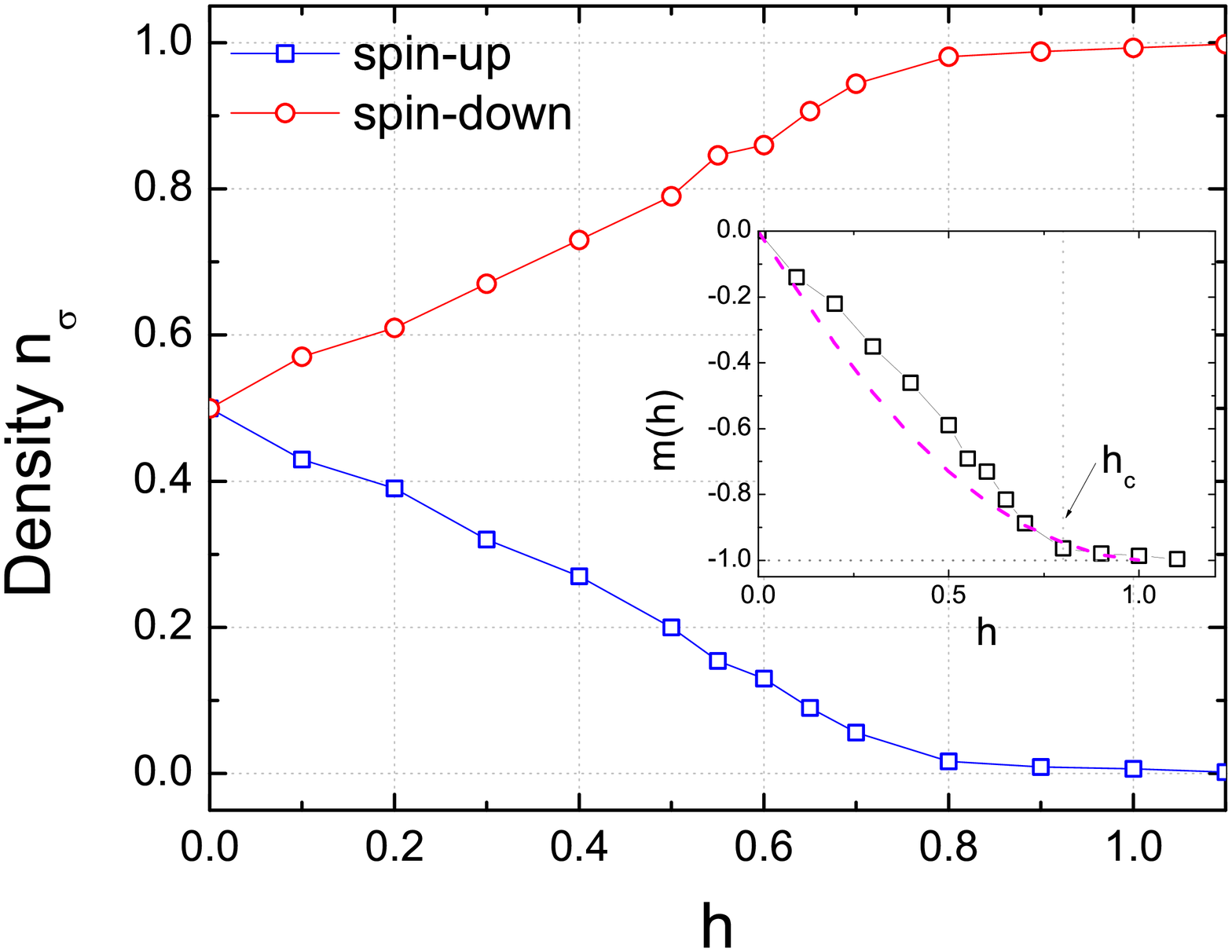}
  \includegraphics[width=0.3\textwidth,angle=0]{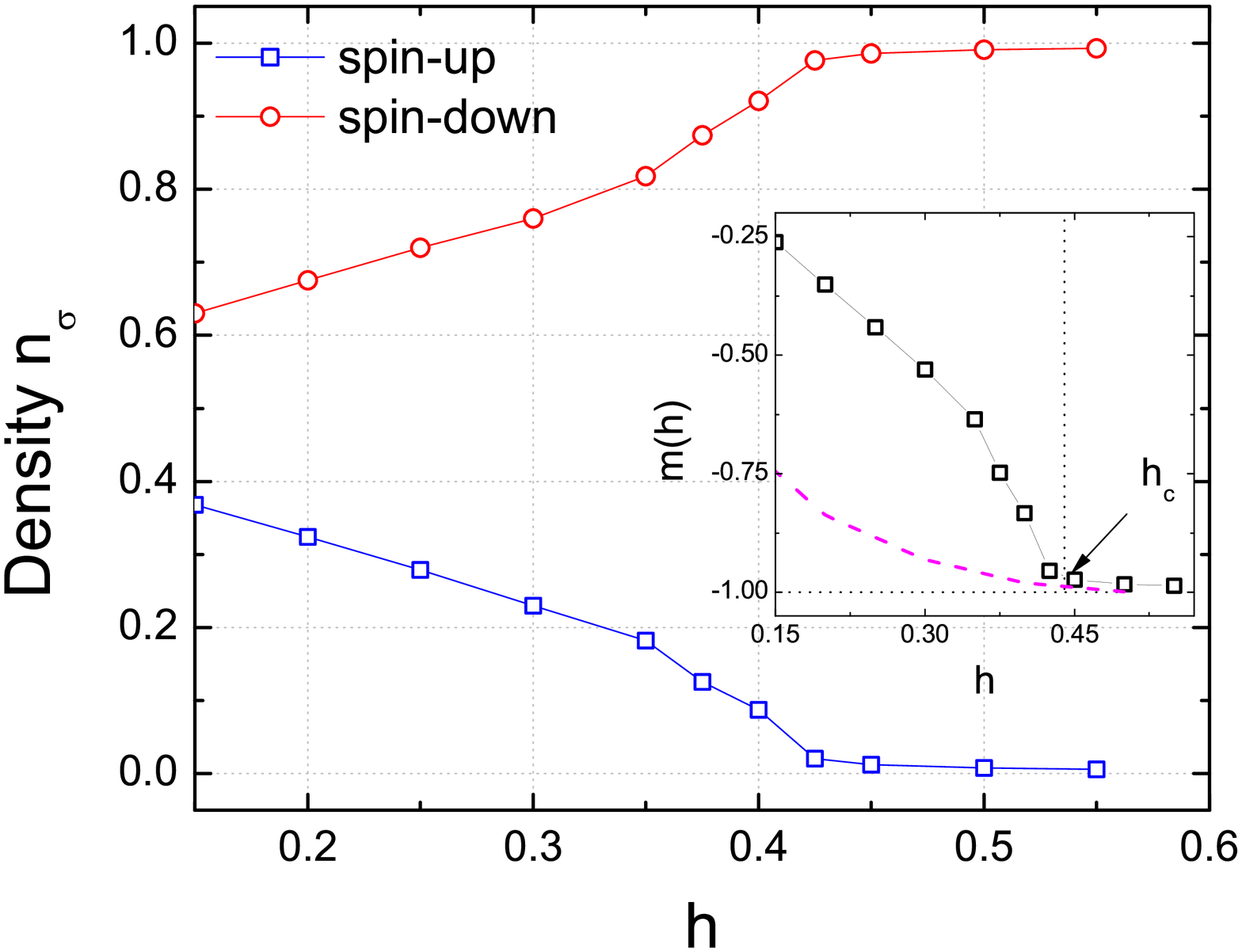}
    \caption{Impurity site spin-dependent occupation density $n_\sigma$ in a magnetic field for interaction strength $U=0.5D=1.0$ (left), $U=D=2.0$ (middle) and $U=1.5D=3.0$ (right), where the occupation density is defined by integrating the spectral density up to Fermi level: $n_\sigma= \int^{0}_{-\infty} d\omega \rho_{\sigma}(\omega)$.
    Inset: The magnetization $m(h)=n_{\uparrow}-n_{\downarrow}$ as a function of magnetic field $h$ (black dotted line).
    The metal-to-insulator transition point is determined to be $h_c$, determined by a kink in magnetization curve and $\vert m(h>h_c)\vert >0.9$.
    The red dashed line shows the mean-field result. }\label{fig:density}
  \end{center}
\end{figure*}

Figure~\ref{fig:density} shows the occupation density at the impurity site as a function of magnetic field strength $h$.
With increasing $h$, occupancy $ n_{\uparrow}$ decreases while $ n_{\downarrow} $ increases (note the definition of exchange field in Eq.~(\ref{eq:ham_hubb})), which effectively reduces  the double
occupancy. To be specific, it is found that the occupation density $ n_{\uparrow(\downarrow)}$ monotonically decreases (increases) with $h$. In the relatively high magnetic field, the system
is close to the fully polarized. Further increasing $h$ will drive a metal-insulator transition near $h_c$, which is determined by
the kink in magnetization curve.

To understand the physics in the regime of weak correlations, the Stoner approximation provides a good description for the field dependent magnetization $m(h)$.~\cite{Levin83,Yoshiro_book}
According to the Hartee-Fock decoupling of the interaction,
\begin{eqnarray*}
&&Un_{i,\uparrow} n_{i,\downarrow} = U (\langle n_{i,\uparrow}\rangle + \delta n_{i,\uparrow}) (\langle n_{i,\downarrow}\rangle + \delta n_{i,\downarrow}) \\
&&\approx U \langle n_{i,\uparrow}\rangle n_{i,\downarrow} + U \langle n_{i,\downarrow}\rangle n_{i,\uparrow}-U \langle n_{i,\uparrow}\rangle \langle n_{i,\downarrow} \rangle \;.
\end{eqnarray*}
Thus the Hamiltonian becomes
\begin{eqnarray}
  \mathcal{H} =  - t \sum_{\langle i,j\rangle, \sigma} c^\dagger_{i,\sigma} c^{\phantom\dagger}_{j,\sigma} +
   \sum_i (h + U \langle n_{i,\downarrow}\rangle -U/2)n_{i,\uparrow}  \nonumber\\
   + \sum_i (-h +U \langle n_{i,\uparrow} \rangle -U/2 )n_{i,\downarrow} \;.
\end{eqnarray}
The corresponding self-consistent equation is
\begin{eqnarray*}
  \langle n_{i,\uparrow} \rangle &=& \int^{0}_{-\infty}d\omega \rho(\omega-h-U \langle n_{i,\downarrow}\rangle +U/2 )\;, \\
  \langle n_{i,\downarrow}\rangle &=& \int^{0}_{-\infty}d\omega \rho(\omega+h -U \langle n_{i,\uparrow}\rangle +U/2 )\;.
\end{eqnarray*}
Using the fact that $ \langle n_{i,\uparrow} \rangle + \langle n_{i,\downarrow}\rangle  =1$, the self-consistent equation reduces to $\langle m_i\rangle = \langle n_{i,\uparrow} \rangle - \langle n_{i,\downarrow} \rangle= \langle m\rangle$:
\begin{equation}\label{}
  \langle m\rangle = (\int^{-h+U \langle m\rangle/2}_{-\infty}d\omega-\int^{h-U \langle m\rangle/2}_{-\infty}d\omega )\rho(\omega ) \;.
\end{equation}

The obtained magnetization curve from the mean-field calculation (red line) is shown in inset of Fig. \ref{fig:density}.
It is found that, in the weakly interacting regime $U<D$,
the calculated magnetization curve agrees well with  the Stoner prediction.
The magnetization exhibits an approximately linear behavior, when the magnetic field strength is weak ($h\leq h_c$),
indicating a nearly constant low-field magnetization susceptibility $\chi(h<h_{c})=m/h$.
In the relatively high magnetic field $h>h_{c}$, the magnetization is  almost flat by increasing magnetic field,
thus the related high-field magnetization susceptibility $\chi(h>h_{c})$ becomes much smaller.

\begin{figure}[b]
  \begin{center}
  \includegraphics[width=0.45\textwidth,angle=0]{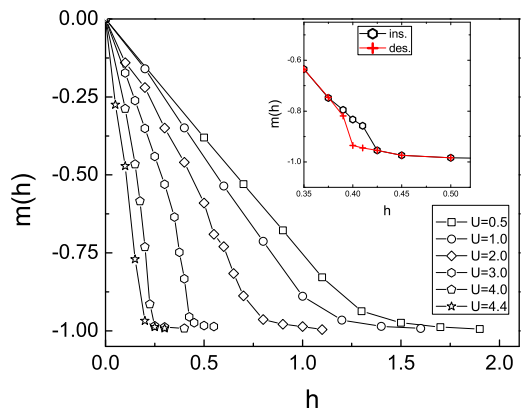}
    \caption{Magnetization as a function of the magnetic field $h$ for various values of the interaction strength $U$.
    In the strong interaction regime, the insulator transition point is determined by a kink in magnetization curve.
    Here the magnetization curve is obtained by sweeping the magnetic field $h$ upward.
    Inset: Hysteresis curve ($U=1.5D=3.0$) by continuously increasing $h$ (black diamond) and by continuously decreasing $h$ (red cross).}
   \label{fig:magnetization}
  \end{center}
\end{figure}

When we tune up the Hubbard interaction strength, the magnetization deviates from the Stoner prediction.
For example, in Fig.~\ref{fig:density}(c) for the magnetization curve at $U=1.5D$
approaching the transition point  $h_c$, the magnetization curve is non-linear and magnetic susceptibility is field-dependent.
To visualize this correlation effect, we plot  the magnetization for various values of the interaction strength in Fig. \ref{fig:magnetization}.
For a stronger interaction, the magnetization is steeper near the phase transition point.
Importantly, a kink in magnetization curve around $h\approx h_c$ indicates the phase transition.
If we continuously tune the magnetic field from low-field to high-field, the magnetization curve is always continuous,
without sudden jump which is also predicted by the Gutzwiller method.~\cite{Vollhardt84}
However, by checking the hysteresis curve, we find a mismatch for increasing  and decreasing fields, as shown in Fig.~\ref{fig:magnetization} (inset).
The hysteresis curve indicates the field-induced transition is of the first-order, consistent with the previous DMFT+ED calculation.~\cite{Laloux94}
The hysteresis curve and related co-existence regime partially support the metamagnetic-like transition in strong interaction regime.
In the weakly interacting regime $U<1.5D=3.0$, the hysteresis vanishes thus the transition there is of the second-order or at least weakly first order.


\subsection{Magnetic susceptibility}
Let us examine the magnetic properties further. Here we calculate the local magnetic susceptibility $\chi_{loc}(\omega)$
of the impurity site defined as
\begin{eqnarray}\label{eq:chi}
  \chi_{loc}(\omega)
  &=& -\langle 0|\hat S^z_{d} \frac{1}{\omega+i\eta-(\hat H-E_0)} \hat S^{z}_{d}|0\rangle \nonumber\\
  &&+ \langle 0|\hat S^z_{d} \frac{1}{\omega+i\eta-(E_0-\hat H)} \hat S^z_{d}|0\rangle ,
\end{eqnarray}
where $\hat S^z_{d}=(n_{d,\uparrow}-n_{d,\downarrow})/2$ is the $z$-component of spin operator at the
impurity site. Physically, the real part of magnetic susceptibility reflects the
slope of magnetization curve, while the imaginary part of magnetic susceptibility is the spin fluctuations,
which is related to the energy absorption or loss  according to the fluctuation-dissipation theorem.

\begin{figure}[b]
  \includegraphics[width=0.95\columnwidth,angle=0]{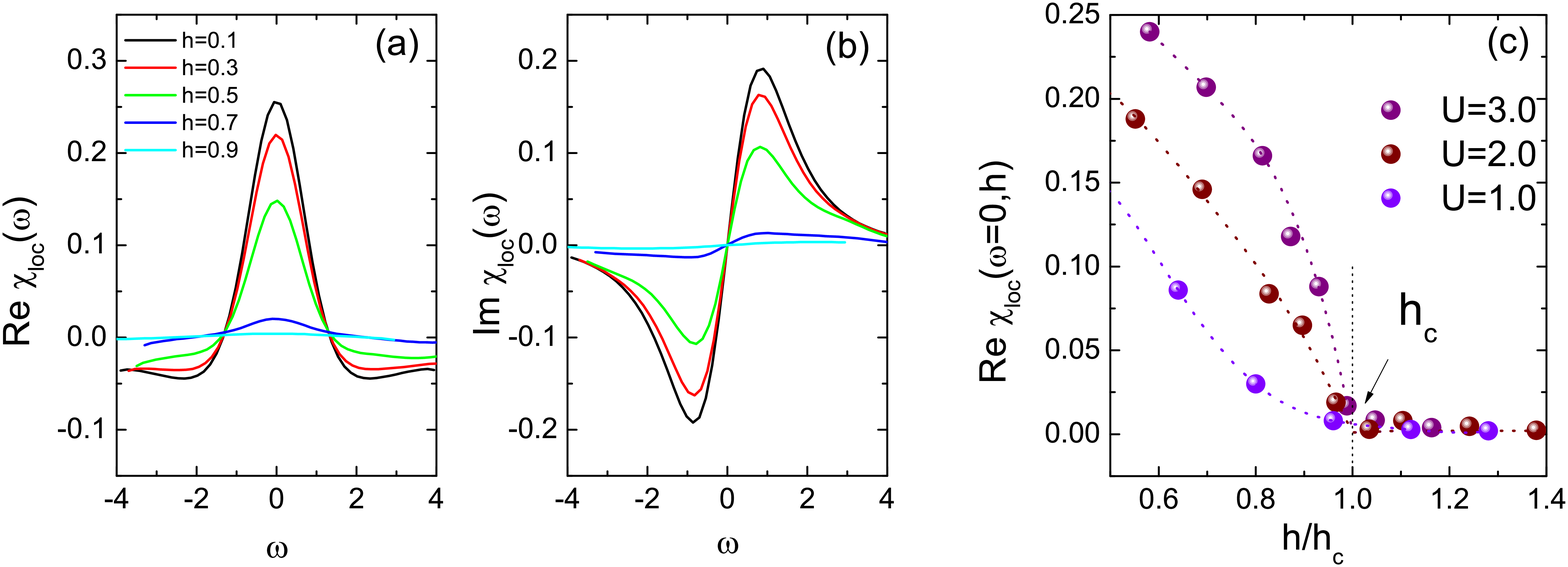}
    \caption{Local magnetic susceptibility at the impurity site calculated for various magnetic field $h$ by setting $U=D=2.0$: (a) Real part $\Re \chi_{loc}(\omega)$ and (b) Imaginary part $\Im \chi_{loc}(\omega)$. (c) $\Re \chi_{loc}(\omega=0)$ as a function of $h$. Dotted line is the best fit to guide for eyes. The phase transition point is determined by the kink around $h_c$.
    }\label{fig:susceptibility}
\end{figure}

We show the calculated magnetic susceptibility $ \chi_{loc}(\omega)$ in Fig. \ref{fig:susceptibility}.
For the real part of susceptibility, we find several features signaling the phase transition.
First, for a given frequency $\omega$, the absolute value of $\chi_{loc}(\omega)$ monotonically decreases with
the increase of magnetic field. Second, $\Re \chi_{loc}(\omega=0)$, the static susceptibility directly relating to
magnetization curve, shows a kink around $h_c$ (Fig. \ref{fig:susceptibility}(c)). In $h>h_c$, the ground state is approximately fully polarized, leading to a
is vanishingly small magnetic susceptibility. The kink around $h\approx h_c$ provides a way to define the phase boundary
between metallic phase and band insulating phase. Third, before entering insulating phase $h<h_c$, it is found that,
the magnetic susceptibility curve
becomes steeper for increased Hubbard interaction when approaching $h_c$ (Fig. \ref{fig:susceptibility}(c)).
This is against the prediction of the Stoner description.~\cite{Laloux94}
Instead, the current observation supports that phase transition follows metamagnetic type of transition.~\cite{Laloux94}
Note that, although the metamagnetic transition was first predicted by the Gutzwiller approximation three decades ago,~\cite{Vollhardt84}
we do not  find the jump or discontinuity in the susceptibility.
It suggests that the Gutzwiller approximation would overestimate the ferromagnetism at the low field.
We notice a discrepancy between the magnetic susceptibility $\Re \chi_{loc}(\omega=0)$
and the magnetization curve $m(h)$. In Fig.~\ref{fig:magnetization}, the magnetization curves are directly obtained from the spin-polarized electron density,
where the slope of $m(h)$ is slightly increasing when approaching critical field $h_c$,
while this behavior does not show up in the magnetization susceptibility $\Re \chi_{loc}(\omega=0)$
in Fig.~\ref{fig:susceptibility}(c). This discrepancy comes from the fact that, except in the limit of $h \rightarrow 0$, the two quantities $\Re \chi_{loc}  (\omega=0)$ (a  local quantity) and $\partial m/\partial h$  (a uniform quantity) can rightfully differ when $h$ enters as a control parameter in the model rather than as an infinitesimal probing field \cite{Laloux94}.

For the imaginary part, it is clearly observed
that $\mathfrak{Im} \chi_{loc}(\omega)$ always turns to zero in the limit of $\omega\rightarrow 0$,
which is guaranteed by  the fact that $\mathfrak{Im} \chi_{loc}(\omega)$ should be an odd function of frequency.
Again, in the high-field regime $h> h_c$, it is found that the imaginary part of susceptibility is vanishing small,
indicating that the spin fluctuation is strongly suppressed for a fully spin-polarized state.


\subsection{Quantum phase transition from entanglement characterization}
Next we discuss the quantum phase transition driven by magnetic field. As shown in Fig.~\ref{fig:spec}, spectral density clearly shows
that the high magnetic field drives the system from a metallic phase to a band insulator. To locate the phase transition point,
usually one can use the magnetization curve as shown in Fig.~\ref{fig:density}, where the kink of magnetic susceptibility
separates the low-field regime from the high-field regime and indicates the phase boundary.
Here, we utilize a state-of-the-art method to determine the phase boundary of quantum phase transition, taking advantage of  the benefit of DMRG calculations.
Currently,  there is growing interest on characterizing quantum phase transitions through the quantum entanglement information.~\cite{Haldane08,Pollmann10}
Despite several attempts of applying these quantum entanglement diagnosis on impurity problems,~\cite{Bayat12,Shirakawa2016}
to the best of our knowledge, the implementation of such kind of entanglement measurements in DMFT calculation is still lacking.
Here, for the first time, we provide an example of phase transition determined by quantum entanglement measurements in DMFT calculations.

In the DMRG method, the system is divided into two parts,
thus the wave function is generally represented as
$\left| \psi \right> = \sum_{i}\sum_{j} \psi_{i,j} \left| i \right>_{\rm L} \otimes \left| j \right>_{\rm R} $,
where $\left| i \right>_{\rm L}$ and $\left| j \right>_{\rm R}$ indicate the bases of the left and right blocks, respectively.
The reduced density matrix $\hat{\rho}_{\rm L}$ for the left block is
$(\hat{\rho}_{\rm L})_{i,i^{\prime}} = \sum_{j} \psi_{i,j} \psi_{i^{\prime},j}^{\ast}$
and the eigenvalue  of $\hat\rho_{\rm L}$ is denoted as $\xi_k$.
The eigenvalues should satisfy the sum rule: $\sum_k \xi_k =1$.
Here we introduce two entanglement measurements related to $\xi_k$.
One is von Neumann entanglement entropy defined as $S = -\sum_k \xi_k \ln \xi_k$,
and the other one is entanglement spectrum as $-\ln \xi_k$.~\cite{Haldane08}
Next we consider the left block enclosing impurity site only (as shown in Fig.~\ref{fig:geom}(a)) and right block enclosing electron bath.

\begin{figure}[t]
  \includegraphics[width=0.45\textwidth,angle=0]{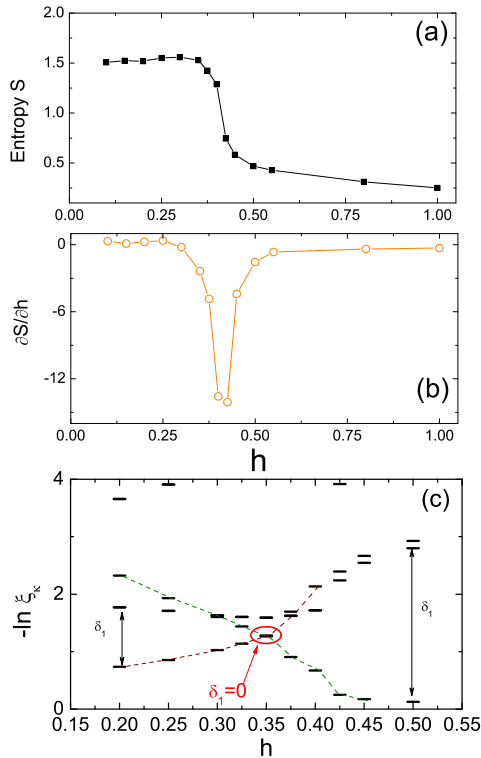}
    \caption{Quantum phase transition determined by the evolution of entropy and entanglement spectrum for $U=1.5D=3.0$.
    (a) Entanglement entropy (black dotted line) for the ground state of the single-impurity Anderson model versus magnetic field $h$.
    (b) Derivative of entropy (orange dotted line), where the dip indicates the phase boundary.
    (c) Low-lying entanglement spectrum for the ground state of the single-impurity Anderson model versus magnetic field $h$.
    We denote $\delta_1$ as the ``entanglement gap'' between the largest eigenvalue and the second largest eigenvalue of the reduced density matrix.
    The red circle marks the entanglement gap closing $\delta_1=0$. The dashed line is the guide for eyes.
    For the entanglement cut, we separate the whole chain into two blocks: left block enclosing the impurity site and one bath site, and the right block enclosing electron bath (see Fig.\ref{fig:geom}(a)).
    }\label{fig:entanglement}
\end{figure}

Figure~\ref{fig:entanglement}(a-b) shows the magnetic field dependence of the entanglement entropy,
and its derivative with respect to the magnetic field, respectively.
The main feature is that the entropy curve exhibits a step-like drop  around critical field $h_c$,
which is a direct evidence for a significant change around the phase boundary in the degree of the quantum entanglement between the impurity site
and electron bath.
Physically, in the band insulating phase,
high magnetic field suppresses the effect of interaction and the fluctuation of the spin, thus
the correlation effect in the fully polarized state is effectively avoided.
Therefore, it is expected that entropy reduces significantly from the metallic phase to the insulating phase by tuning the magnetic field.

Moreover, we observe some more evidence of quantum phase transition from the entanglement spectrum,
as shown in Fig.~\ref{fig:entanglement}(c). In particular, in the vicinity of the phase boundary, the two largest eigenvalues of the reduced density matrix
cross with each other when the magnetic field is increased.
The corresponding ``entanglement gap'' closes ($\delta_1=0$) just before entering the insulating phase, and then reopens.
In the high magnetic field regime, most of the weight of the reduced density matrix is carried by only one eigenvalue (the largest eigenvalue $\max\{\xi_k\}>0.9$),
suggesting that the fully polarized state is actually close to the direct product state with small entanglement correlations.
Finally, we point out that the phase boundaries obtained from different methods agree very well with each other.
For example, for $U=1.5D=3.0$  the critical field $h_{c}\approx 0.4$  obtained from the entropy jump (in Fig. \ref{fig:entanglement}(a)),
is very close to
$h_{c}\approx 0.42$ obtained from the magnetic susceptibility (in Fig. \ref{fig:susceptibility}).
Therefore,  we conclude  confidently  that the entropy change and corresponding level-crossing in the entanglement spectrum directly reveal the quantum phase transition.

\subsection{Quantum phase diagram}
\begin{figure}
  \includegraphics[width=0.35\textwidth,angle=0]{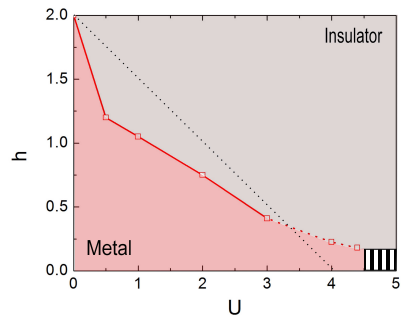}
    \caption{Quantum phase diagram versus interaction strength $U$ and magnetic field $h$.
    The red line represents the phase boundary determined by DMFT+DMRG calculations.
    The black dashed line represents the classical phase boundary: $2D-2h-U=0$, where $2D=4.0$ is the bandwidth of the Bethe lattice.
    The shaded marks the regime without stable solution in current scheme.
    }\label{fig:phase}
\end{figure}

Our systematic analyses presented above enable us to present a quantum phase diagram for the Hamiltonian Eq.~(\ref{eq:ham_hubb}),
as functions of interaction strength $U$ and magnetic field $h$ in Fig.~\ref{fig:phase}.
We find two different phases: a metallic phase and a band insulator phase.
(At $h=0$, we identify a coexistence regime (marked by shadow) starting from $U_c\approx 4.5$, which is consistent with the estimation of  $U_c(h=0)\approx4.76$ in Refs.~\onlinecite{Raas08} and \onlinecite{Capone2007})
When both $h$ and $U$ are small, the ground state is metallic but with finite magnetization $m\neq0$.
By increasing $h$, the magnetic field drives the system into a band insulator phase with
spin being fully polarized.
The phase boundary is determined by the kink of magnetization curve.
We find this boundary is very close to the one  determined by the entanglement entropy, so
we do not distinguish these two phase boundaries.
The dashed line in Fig.~\ref{fig:phase} is the classical phase boundary $U+2h-2D=0$.
Overall, in Fig. \ref{fig:phase}, we find that the stronger the Hubbard interaction, the weaker the critical magnetic field is for the quantum phase transition.
Physically, stronger interaction tends to quench the kinetic degrees of freedom so that a weaker magnetic field
is sufficient to split spin-up and spin-down bands.  By comparing the classical and quantum phase boundaries,
it is found that classical method overestimates the phase boundary for the weak interaction ($U<3.2$), while the quantum critical field $h_c$ is higher than classical estimation in the regime $3.2<U<4.4$.

In addition, in the vicinity of the Mott insulator phase ($U>4.4$), we identify a band insulator phase by tuning magnetic field $h>0.2$,
however, we do not find a convergent solution within DMFT scheme for small magnetic field $h<0.2$ (as shown in the shaded area near the right corner at the bottom of the phase diagram).
This can be understood by the strong interaction limit of Hubbard model at half-filling. That is, the effective Hamiltonian reduces to
the Heisenberg model with only spin degree of freedom frozen on each local site, which reads $H=\sum_{<ij>} J_{ij} S_i \cdot S_j$ with $J_{ij}=J\sim4t^2/U$.
Therefore, in the absence of a magnetic field, the system intrinsically favors an antiferromagnetic state,
 which is beyond the single-site DMFT study without breaking the original lattice translational invariance.~\cite{Hweson07}
Note that, in Fig.~\ref{fig:phase}, the metal-to-insulator phase boundary is determined by sweeping the magnetic field $h$ upward.
We caution here that  the phase boundary shown here is  the up-limit of the metallic phase.
Although we identify a hysteresis curve near $U\approx3.0$ (as shown inset of Fig. \ref{fig:magnetization}),
the coexistence region is tiny and we do not show it in Fig. \ref{fig:phase}.

\section{Concluding Remarks}
We have studied the magnetic field driven metal-to-insulator transition in a half-filled Hubbard model on the Bethe lattice (or in the limit of infinite dimensions).
To do so, we have developed a scheme within the DMFT by solving the impurity model by  means of DMRG.
First, the high-resolution field-dependent spectral density shows that the Kondo resonance peak splits in the weak magnetic field;
while in high magnetic field, the spin-up and spin-down  bands move away from the
Fermi level and finally form a spin polarized band insulator.
Second, in weak interaction regime, we have identified a smooth crossover from paramagnetic metal to the fully polarized band insulator,
with magnetization continuously increasing to unity. In the strong interaction regime, hysteresis curve indicates
a metamagnetic-like phase transition despite the coexistence regime being very  small.
Third, the phase boundary has been determined by two different methods. One is the kink in magnetic susceptibility, which
separates the low-field regime from the high-field regime. The other one is the step-like jump  of the entanglement entropy and corresponding
entanglement gap closing, which reveal distinct quantum entanglements between impurity site and electron bath in two different phases.

Experimentally, the  liquid $^3$He, which is regarded as a canonical Landau Fermi liquid,
is believed to be a testbed for the field-driven metal-to-insulator transition.
However, earlier study~\cite{Wiegers91} found a smooth variation of the magnetization with the applied field, instead of  a metamagnetic-like transition,
which is more compatible with the Stoner's description.~\cite{Levin83,Yoshiro_book}
Our study has shown that the metamagnetic-like transition only occurs in the strong interaction regime.
The magnetization is always continuous when one increases magnetic field upward.
One reasonable explanation for previous $^3$He experiment~\cite{Wiegers91} is that
the effective interaction strength is less than the estimate made in earlier work.~\cite{Vollhardt84}
To settle down this controversy, we suggest a hysteresis curve measurement in liquid $^3$He.
Besides the liquid $^3$He, many attempts have been made to search for metamagnetism in different systems,
including for example,  
quasi-two-dimensional organic conductor $\kappa$-(BEDT-TTF)$_2$Cu[N(CN)$_2$]Cl.~\cite{Kagawa04}
To sum up, our present work suggests to look for experimental systems, which can be reasonably modeled by
the single-band Hubbard model, with its effective interaction strength tunable from weak to strong regimes.
Considering the recent rapid-development in optical lattice, we expect that the cold-atom systems
would serve as an ideal playground to realize the Stoner-like and metamagentic-like transition in laboratory.

\acknowledgments
WZ thanks for Carsten Raas and Eric Jeckelmann for simulating discussion. 
We also thank for the helpful comments from the reviewers. 
This work was supported by DOE National Nuclear Security Administration through Los Alamos National
Laboratory LDRD Program (W.Z. \& J.-X.Z.), the National Science Foundation  grant DMR-1408560  (D.N.S.), and in part by the Center for Integrated Nanotechnologies, a U.S. DOE Basic Energy Sciences user facility.


\clearpage
\begin{widetext}
\widetext

\appendix
\begin{appendices}

\section{DMFT self-consistent condition for the Bethe lattice}
\label{sec:dmft_bethe}
In this Appendix, we briefly discuss the self-consistent condition, hybridization function for the Bethe lattice, with or without magnetic field.
The Bethe lattice is interesting due to its specific form of density of states (DOS),
which can simplify the DMFT self-consistent condition.
It enables us not to use any other
feature from the Bethe lattice other than the DOS form.

\subsection{Zero magnetic field}
On the Bethe lattice, the single-particle DOS (in the absence of interaction term) takes a semi-elliptic form:
\begin{equation}
  \rho^0(\omega) = -\frac{1}{\pi}\mathfrak{Im}G^0(\omega)=\frac{2}{\pi D^{2}}\sqrt{D^{2}-\omega^{2}},
\end{equation}
where $2D$ stands for the band width of system and  the
bare lattice Green's function $G^0(\omega)$ takes a particularly simple
continued fraction representation with constant coefficients
\begin{equation}
\label{eq:bethe_contfrac}
G^0(\omega) =
  \frac{1}{\omega-
    {\displaystyle\frac{\frac{D^2}{4}}{\omega-
        {\displaystyle\frac{\frac{D^2}{4}}{\omega-\cdots}}}}}.
\end{equation}
With the help of the Dyson equation, the full lattice Green's function $G(\omega)$ can be expressed to be
\begin{equation}
\label{eq:dyson1}
\frac{1}{G(\omega)} = \frac{1}{G^0(\omega-\Sigma(\omega))} =\omega -\Sigma(\omega) -
 \frac{\frac{D^2}{4}}{\omega-\Sigma -
    {\displaystyle\frac{\frac{D^2}{4}}{\omega-\Sigma -
        {\displaystyle\frac{\frac{D^2}{4}}{\omega-\Sigma\cdots}}}}} 
 = \omega -\Sigma(\omega) - \left(\frac{D^2}{4} \right) G(\omega),
\end{equation}
where $\Sigma(\omega)$ is the self-energy function.
Importantly,  several remarks are in order. First, we have assumed that self-energy function $\Sigma(\omega)$
is uniform in real-space thus it is independent of momentum quantum number, which is one key assumption of DMFT.
Due to this assumption, self-energy function behaves as a global energy shift to frequency $\omega$.
Last but not least, the continued fraction does not change
when it is evaluated at a deeper level because its coefficients are constant.

On the other hand, we can write the bare Green's function
of the Anderson impurity model $g^0(\omega)$ with the help of the so-called hybridization function
$\Gamma(\omega)$ as
\begin{equation}
g^0(\omega) = \frac{1}{\omega-\Gamma(\omega)}
\end{equation}
where the continued fraction of $\Gamma(\omega)$ is
\begin{equation}
  \label{eq:hyb_coef}
  \Gamma(\omega) =
  \frac{V^2}{\omega-\varepsilon_0-
    {\displaystyle\frac{\gamma_0^2}{\omega-\varepsilon_1-
        {\displaystyle\frac{\gamma_1^2}{\omega-\cdots}}}}}.
\end{equation}
For an infinite homogeneous system we have  $\gamma_i = D/2$, $\varepsilon_i=0$ and $V = D/2$.
From the Dyson equation,
the impurity Green's function of Anderson impurity model  reads
\begin{equation}
\label{eq:dyson2}
\frac{1}{g(\omega)} =\frac{1}{g^0(\omega)}-\Sigma(\omega) =\omega- \Sigma(\omega) -\Gamma(\omega).
\end{equation}

Based on the self-consistency condition (\ref{eq:self-con}), we
set Eqs. (\ref{eq:dyson1},\ref{eq:dyson2}) equal and obtain the simpler
self-consistency condition:
\begin{equation}
  \label{eq:hyb_green_sc}
  \Gamma(\omega) = \frac{D^{2}}{4}G(\omega).
\end{equation}
This equation is simple and it provides a direct way to compute
the hybridization function $\Gamma(\omega)$ (Eq. (\ref{eq:hyb_coef})) of the next iteration of
the Anderson impurity model from the lattice propagator $G(\omega)$.

\subsection{Nonzero magnetic field}
We now discuss the case in the presence of magnetic field $h$ (exchange field in spin space).
In the non-interacting limit $U=0$, under the effect of magnetic field $h$,
it is easy to see the form of single-particle spin resolved DOS $\rho^0_{\sigma}(\omega) :=-\frac{1}{\pi}\mathfrak{Im}G^0_{\sigma}(\omega)$ as
\begin{equation}
  \rho^0_{\uparrow}(\omega) = \frac{2}{\pi D^{2}}\sqrt{D^{2}-(\omega-h)^{2}}, \rho^0_{\downarrow}(\omega) = \frac{2}{\pi D^{2}}\sqrt{D^{2}-(\omega+h)^{2}},
\end{equation}
where the semi-elliptic DOS has the relation to Eq. (\ref{eq:bethe_dos}) as $\rho^0_\sigma(\omega)=\rho^0(\omega+\sigma h)$, and
they host a symmetry relation as:
\begin{equation}
  \rho^0_{-\sigma}(\omega) =  \rho^0_{\sigma}(-\omega) 
\end{equation}

The related spin resolved Green's function $G^0_\sigma$ with semi-elliptic
$\rho^0_\sigma(\omega)$ can be represented, similar to Eq. (\ref{eq:bethe_contfrac}):
\begin{equation}
G^0_\sigma(\omega) =
  \frac{1}{\omega+\sigma h-
    {\displaystyle\frac{\frac{D^2}{4}}{\omega+\sigma h-
        {\displaystyle\frac{\frac{D^2}{4}}{\omega+\sigma h-\cdots}}}}} = G^0(\omega+\sigma h).
\end{equation}
Via the Dyson equation and introducing related self-energy function $\Sigma_\sigma(\omega)$,
 the full Green's function the lattice can be expressed to be
\begin{eqnarray}
\nonumber
\frac{1}{G_\sigma(\omega)} &=& \frac{1}{G_\sigma^0(\omega-\Sigma_\sigma)} =\omega+\sigma h -\Sigma_\sigma -
 \frac{\frac{D^2}{4}}{\omega+\sigma h-\Sigma_\sigma -
    {\displaystyle\frac{\frac{D^2}{4}}{\omega+\sigma h-\Sigma_\sigma -
        {\displaystyle\frac{\frac{D^2}{4}}{\omega+\sigma h-\Sigma_\sigma\cdots}}}}}
\\
 &=& \omega+\sigma h -\Sigma_\sigma(\omega) - \left(\frac{D^2}{4} \right) G_\sigma(\omega),
\label{eq:spin-compare2}
\end{eqnarray}
where we explicitly show that the continued fraction does not change
when it is evaluated at a deeper level because its coefficients
are constant.

On the other hand, the bare Green's function
of the Anderson impurity model becomes
\begin{equation}
g^0_\sigma(\omega) = \frac{1}{\omega+\sigma h-\Gamma_\sigma(\omega)} = \frac{1}{\omega+\sigma h-\Gamma(\omega+\sigma h)} =  g^0(\omega+\sigma h)
\end{equation}
where the spin resolved hybridization function $\Gamma_\sigma(\omega)$ as
\begin{equation}
  \Gamma_\sigma(\omega) =  \Gamma(\omega+\sigma h)=
  \frac{V^2}{\omega+\sigma h-
    {\displaystyle\frac{\gamma_0^2}{\omega+\sigma h-
        {\displaystyle\frac{\gamma_1^2}{\omega+\sigma h-\cdots}}}}}.
\end{equation}
Again, using Dyson equation,
the full Green's function of the Anderson impurity model reads
\begin{equation}
\frac{1}{g_\sigma(\omega)} = \omega+\sigma h -\Gamma_\sigma(\omega) - \Sigma_\sigma(\omega).
\label{eq:spin-compare1}
\end{equation}

Based on the self-consistency condition (\ref{eq:self-con}), we
set Eqs. (\ref{eq:spin-compare1},\ref{eq:spin-compare2}) equal and obtain the simpler
self-consistency condition
\begin{equation}
  \Gamma_\sigma(\omega) = \frac{D^{2}}{4}G_\sigma(\omega).
\end{equation}
This was first derived by Ref. \onlinecite{Laloux94}, and is  similar to Eq. (\ref{eq:hyb_green_sc}), except that , the spin-up and spin-down part
are not equivalent due to the presence of external magnetic field. Note that the spin-dependent Green's function satisfy the following relation:
\begin{equation}
 G_{-\sigma}(\omega) = G_{\sigma}(-\omega) \;.
\end{equation}

\section{DMRG solution for single impurity Anderson model without magnetic field}
\label{sec:SIAM_test1}
In this Appendix, to numerically verify the two-component mapping scheme,
we apply the dynamical DMRG technique to the single impurity Anderson model at half-filling,
which provides a very good benchmark and testing ground. The usual single impurity Anderson model
is written as
\begin{equation}
\label{eq:ham_siam}
  H = U(n_{d,\uparrow}-1/2) (n_{d,\downarrow}-1/2)+ \sum_{\sigma} V_{\sigma} (d^{\dagger}_{\sigma} c_{1,\sigma} +h.c.) + \sum_{i=1,\sigma}^{L-1} \varepsilon_i c^{\dagger}_{i,\sigma} c_{i,\sigma} + \sum_{i,\sigma} \gamma_i(c^{\dagger}_{i,\sigma} c_{i+1,\sigma}+h.c.),
\end{equation}
where the coefficient of bath electrons comes from the hybridization function $\Gamma(\omega)$ Eq. (\ref{eq:hyb_coef}) (or DOS $\rho_0(\omega)$ given by Eq. (\ref{eq:bethe_dos})) of the Bethe lattice: $\varepsilon_i=0$, $\gamma_i=D/2$ and $V_{\sigma}=D/2$ ($2D$ is band-width of Bethe lattice). The $d$ electron represents impurity that
is correlated due to the repulsive interaction $U > 0$.

We are interested in the dynamical properties relating to the one-particle impurity Green function:
\begin{equation}
	G_{\sigma}(\omega) =\lim_{\eta\rightarrow 0^{+}} \langle 0| \hat d_\sigma^{\dagger}  \frac{1}{\hat H- E_0+\omega-i\eta} \hat d_\sigma |0\rangle +
                         \langle 0| \hat d_\sigma  \frac{1}{E_0-\hat H+\omega+i\eta} \hat d_\sigma^\dagger |0\rangle,
\end{equation}
where $|0\rangle$ stands for the ground state and $E_0$ is ground state energy. The spectral density
is therefore obtained by $\rho(\omega)=-\frac{1}{\pi} \Im G(\omega)$.

Here we show the spectral density $\rho(\omega)$ and real part of Green function $G_{\sigma}(\omega)$
for Hamiltonian, Eq. (\ref{eq:ham_siam}), in Fig.~\ref{sfig:1dbethe}.
Without magnetic field (or any other mechanism breaking symmetry between spin-up and spin-down),
the Green function has no dependence on the spin index $\sigma$.
For finite size calculation, we choose a chain with $L=80$ fermionic sites
(after two-component mapping in Sec. \ref{sec:mapping}, we actually work on a chain with $L'=160$ spinless fermion sites).
We conclude that our calculation can recover all features known for this model.~\cite{Raas04}
First, in the absence of interaction $U=0$, the spectral density shows a semielliptic form,
which almost repeat the result from continuous version of the single impurity Anderson model.
Second, it is found that spectral density is pinned to $\rho(\omega=0)=\frac{2}{D\pi}$
which is requirement from Friedel sum rule.~\cite{Anders91} This fact serves as a convincing evidence for the reliability
of our numerical algorithm.
Third, for the Kondo resonance peak around the Fermi surface,
the half width of central Kondo resonance peak is the rapidly narrowing by increasing the interaction strength $U$.
This behavior is also consistent with the expectation that the Kondo energy scale (Kondo temperature)
monotonically decreases with interaction strength.
Fourth, when interaction strength is larger than band width $U\geq 2D$, two symmetric non-coherent peaks (Hubburd satellites)
develop in the high frequency regime.
The non-coherent peak structure proves the great advantage of DMRG, compared to normal NRG calculations:
Since the low-frequency and high-frequency regime are deal with equivalently in DMRG, both Kondo resonance
and non-coherent peaks can be fairly viewed.
We also compared our results with the recent publications using DMRG-based techniques,
and the results are consistent with the publications as well.~\cite{Raas08}

\begin{figure}[!htb]
    \includegraphics[width=0.45\columnwidth,angle=0]{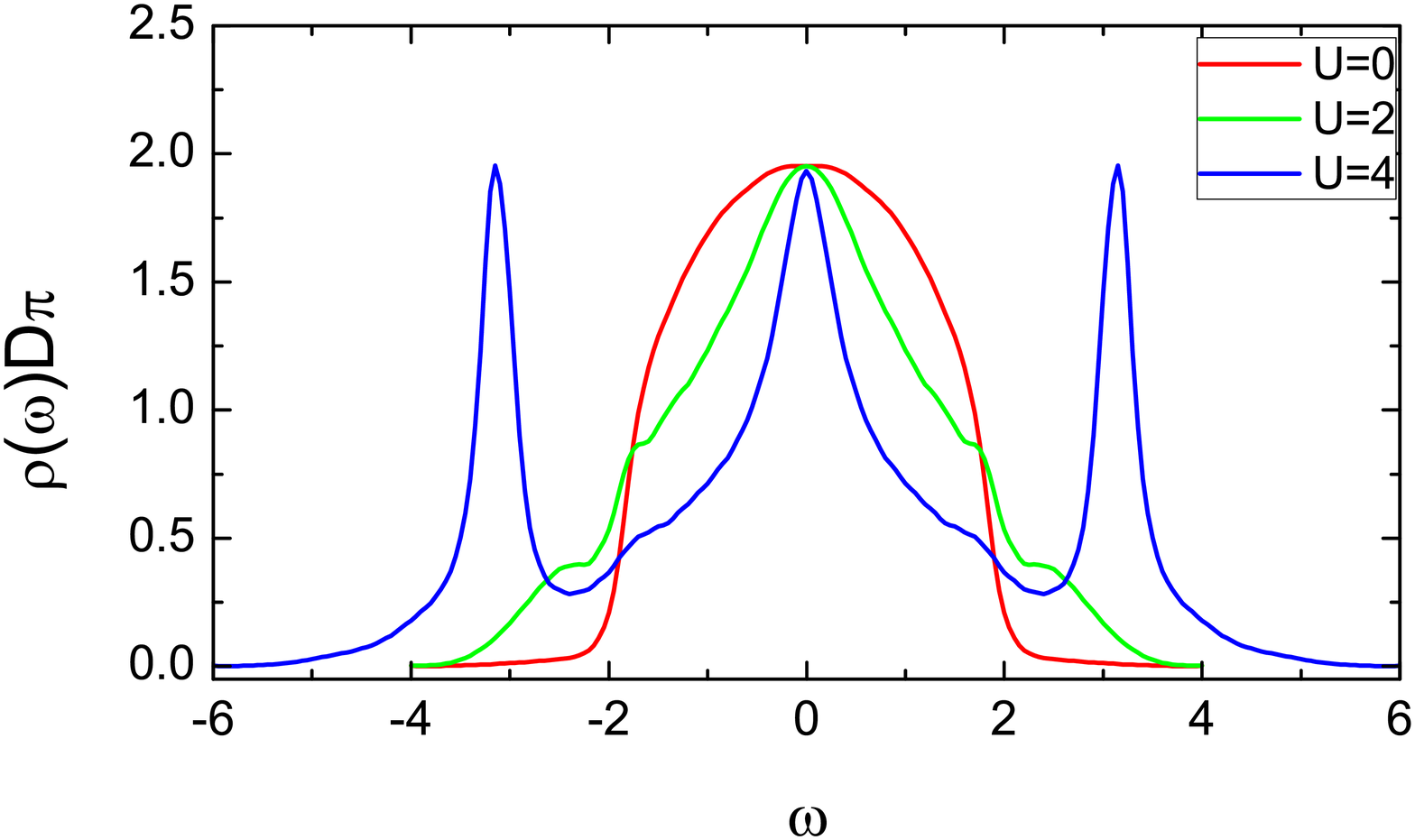}\\
    \includegraphics[width=0.45\columnwidth,angle=0]{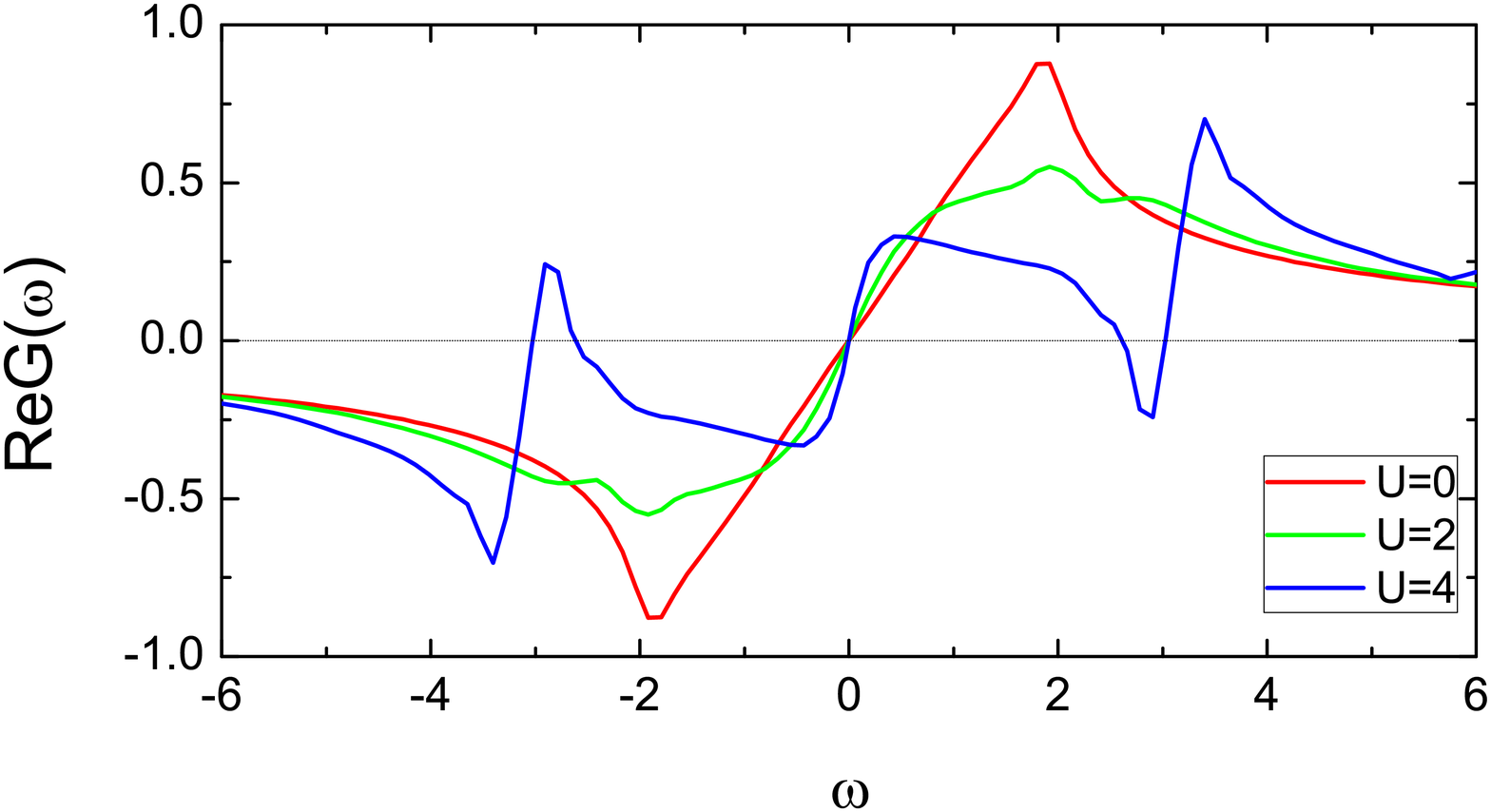}
    \caption{Dynamical DMRG solution for one dimensional single-impurity Anderson model (by setting $V=D/2$, $\gamma_i=\gamma=1.0$ in Eq. \ref{eq:hyb_coef}) for different interaction strength $U=0,D,2D$:
    (Left) Spectral densities (imaginary part of Green function by scaling a global constant $\pi D$) and (Right) Real part of Green functions as a function frequency $\omega$.
     For DMRG simulation, we choose a chain length $L=80$ fermionic sites (after mapping into two-component spinless model, we have $L'=160$ lattice sites).
    We kept $128$ states in each DMRG block and the resulting projection error is less than $10^{-10}$. In the dynamical DMRG calculation, we use a smearing energy $\eta=0.1D$ before deconvolution calculation.
    }\label{sfig:1dbethe}
\end{figure}

\section{DMFT+DMRG solution of single-orbital Hubbard model on the Bethe lattice without magnetic field}
\label{sec:SIAM_test2}
In this Appendix, we solve a single-orbital Hubbard model on the Bethe lattice in the absence of magnetic field, which is given by the Hamiltonian
\begin{equation}
  \mathcal{H} =
  U \sum_i \left(n_{i,\uparrow  }-\frac{1}{2}\right)
           \left(n_{i,\downarrow}-\frac{1}{2}\right)
  - t \sum_{\langle i,j\rangle, \sigma}
    c^\dagger_{i,\sigma} c^{\phantom\dagger}_{j,\sigma}
\end{equation}
where $c^\dagger_{i,\sigma}$ creates one electron with spin$-\sigma$ at site $i$  and
$n_{i,\sigma}$ is occupation operator.
The basic physics of the hubbard model comes from
the competition between the local repulsive interaction and kinetic term consisting
of hopping from one site to the other site. The interaction is diagonal in real space and
hence tends to make the electrons local in real space,
while the kinetic energy is diagonal in momentum space and hence tends to make
the electrons extended in real space. So the interaction favors an insulating phase, whereas the
kinetic energy favors a metallic phase, depending on the relative strength of $U/t$.

Here we solve the Hamiltonian, Eq.~(\ref{eq:ham_hubb}), with DMFT scheme. The related key DMFT self-consistent
condition has been discussed in Sec.~\ref{sec:dmft_bethe}. Fig.~\ref{sfig:2dbethe} shows our results for
various interaction strength $U$ in the metallic phase ($U<U_c\approx 2.6 D$).
Here we choose the one-dimensional impurity model enclosing $L=80$ fermionic sites,
which is solved by DMRG algorithm by limiting each DMRG block with dimension $M=128$.
The obtained projection error in DMRG calculation are all negligible small (less than $10^{-10}$),
indicating good convergence of DMRG output from core impurity model.
As to computational performance, by setting parameter $U=2D$,
the typical (physical) time cost is $63$ minutes for each DMFT loop (on two $3.90$GHz cores). Here
we set the simulation parameter as the broadening energy $\eta=0.1D$ and frequency scan step $\Delta \omega=\eta$ ($\omega\in [-6.0,6.0]$),
and use the mixed Bath discretization (see Appendix \ref{sec:discre}). For more details about computational performance, please see Appendix.~\ref{sec:time}.

The obtained spectral densities faithfully recover the previous DMFT+DMRG calculations,~\cite{Raas08}
with key features including the pinning criterion $\rho(\omega=0)=\frac{2}{\pi D}$ for all interaction strength, and
the side peaks at the inner edges of Hubbard bands in strong interaction regime ($U=2D$). Compared with
previous numerical renormalization group calculations, current DMFT+DMRG scheme deal with low frequency
and high frequency with the equal weight, thus we can get correct both Kondo resonance peak in the low frequency
and Hubbard satellite bands (non-coherent peak) in the high frequency.
Compared with the Chebyshev-based simulations, current DMFT+DMRG reaches a better convergence
(In Ref.~\onlinecite{Wolf14}, the pinning criterion
violates when interaction strength becomes strong or simulation system size increases larger than $L=80$.
The authors argued that the linear prediction overestimates
the height of the central Kondo peak (see Appendix in Ref.~\onlinecite{Wolf14}).
We didnot observe these drawbacks in our current DMFT+DMRG realization).
In a word, under DMFT+DMRG scheme, by using the two-component mapping, a better convergence and
computational performance is available.

\begin{figure}[!htb]
    \includegraphics[width=0.9\columnwidth,angle=0]{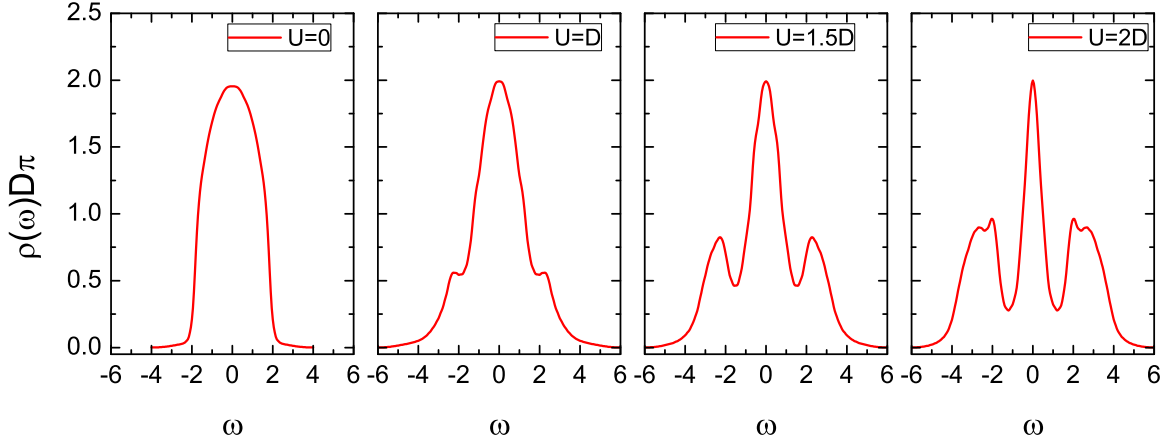}
    \caption{Spectral densities of single-orbital Hubbard model on the Bethe lattice obtained by DMFT scheme.
    We choose the impurity model enclosing $L=80$ fermionic sites, which is solved by DMRG algorithm by limiting each DMRG block with dimension $M=128$.
    Before the deconvolution calculation, we select the broadening parameter as $\eta=0.1D$.}\label{sfig:2dbethe}
\end{figure}

\section{Bath discretization scheme}
\label{sec:discre}
Since the numerical calculations are performed on the lattice system,
we have to discretize the continuous bath band or
continuous hybridization function and construct lattice model.
Here we discuss different bath discretization schemes for DMFT calculation:
a linear discretization, a logarithmic discretization and a hybridization discretization.

When we have a continuous bath band (or DOS function $\rho(\omega)$), it can be proved that
the effective coupling between a single impurity site and continuous bath is
reproduced exactly by the following Hamiltonian:
\begin{eqnarray}\label{eq:hw}
 H_{imp-bath} &=&
\sum_{\sigma = \uparrow, \downarrow}\int {\rm d}\omega \omega c^\dag_{\omega,\sigma}c_{\omega,\sigma}
+V \sum_{\sigma = \uparrow,\downarrow} \int {\rm d}\omega  \sqrt{\rho (\omega)} \left( d^\dag_{\sigma}c_{\omega,\sigma}+{\rm H.c.}\right),
\end{eqnarray}
where $c^\dag_{\omega,\sigma}$ ($c_{\omega,\sigma}$) is the creation (annihilation)
operator of a bath electron which represents the eigenstate with
energy $\omega$ and spin $\sigma$.

Next, we discretize the energy $\omega$ with three ways:
\begin{itemize}
  \item Linear discretization.---  $\{\omega_m\}=\omega_{min}+\Delta\omega m$, where $\Delta=(\omega_{max}-\omega_{min})/L$ is the energy step.
  \item Logarithmic discretization.--- $\{\omega_m\}=\{\omega^+_m\}\oplus\{\omega^{-}_m\} ,\,\,\,\, \omega_m^{\pm} = \pm D \Lambda^{-m}$, where $2D$ is the band width, $\Lambda\,(>1)$ is a parameter which sets a series of intervals
in $\omega_m^{\pm}$'s with $m=0,1,\dots,M-1$, and we set $\omega_{M}^{\pm} = 0$.
This is the original scheme applied to numerical renormalization group introduced by Wilson.~\cite{Wilson75}
  \item Mixed discretization.--- $\{\omega_m\}=\{\omega^I_m\}\oplus\{\omega^{O}_m\}$ ,\,\,\,\, $\omega_m^{I} = \pm D \Lambda^{-m}$ for $|\omega_m|<\omega^{hyb}$,\,\,\,\,$\{\omega_m^O\}=\omega_{min}+\Delta\omega m$ for $|\omega|>\omega^{hyb}$. This is actually a combination of linear and logarithmic discretization scheme.
\end{itemize}

Defining
a representative fermion operator $c^{\dagger}_{m,\sigma}$
for each energy interval  $[\omega_{m-1},\omega_{m}]$,
the coupling between impurity and bath can now be expressed as
\begin{eqnarray}\label{eq:hr}
 H_{imp-bath} &=&
\sum_{m} \sum_{\sigma = \uparrow,\downarrow}  \xi_{m} c_{m,\sigma}^{\dagger} c_{m,\sigma}
 + \sum_{m}\sum_{\sigma = \uparrow, \downarrow} \mu_{m}
d_{\sigma}^{\dagger} c_{m,\sigma} + {\rm H.c.},
\end{eqnarray}
where
\begin{eqnarray}
 \mu_m  = V \left[ \int_{\omega_{m-1}}^{\omega_{m}} {\rm d}\omega \rho (\omega)\right]^{1/2},\,\,\,\,\,
 \xi_{m}= \frac{\int_{\omega_{m-1}}^{\omega_{m}} {\rm d}\omega \rho (\omega) \omega }
 {\int_{\omega_{m}}^{\omega_{m-1}} {\rm d}\omega \rho (\omega) } .
\end{eqnarray}

The last step is to to
map the preceding Hamiltonian on a chain Hamiltonian with only nearest neighbor hoppings ($\varepsilon_i,\gamma_i$ in main text)
by using the Lanczos algorithm. This step is the same with the one employed in the numerical renormailzation group method.~\cite{Bulla08}

The logarithmic discretization scheme has much denser energy
meshes in low frequency regime, but has much less energy meshes for
high frequency regime.
Therefore, the logarithmic discretization scheme cannot capture the properites in high-energy scales
with a high accuracy.
In contrast, the linear discretization scheme distributes
the energy meshes equally for all frequency scales and deal with all frequency scales with equal weight.
Unfortunately, linear discretization scheme usually takes much longer time to converge in DMFT calculation.
One balanced way is to use the mixed (hybridization) scheme. We use logarithmic discretization in low energy regime while
the linear discretization in high frequency regime, which gives reliable results in both low and high frequency regime.
As shown one example in Fig. \ref{sfig:compdisc} (left), for logarithmic discretization scheme,
the spectral density shows unexpected fluctuation in high frequency regime (or logarithmic discretization doesnot reach convergence within the same parameter setting). However, for mixed discretization scheme (right),
we can reach a smooth curve in all frequency regime.
Here, we conclude that mixed scheme shows better performance than the other two schemes, thus in this paper
we used mixed scheme throughout.

\begin{figure}[!htb]
  \begin{center}
    \includegraphics[width=0.4\columnwidth,angle=0]{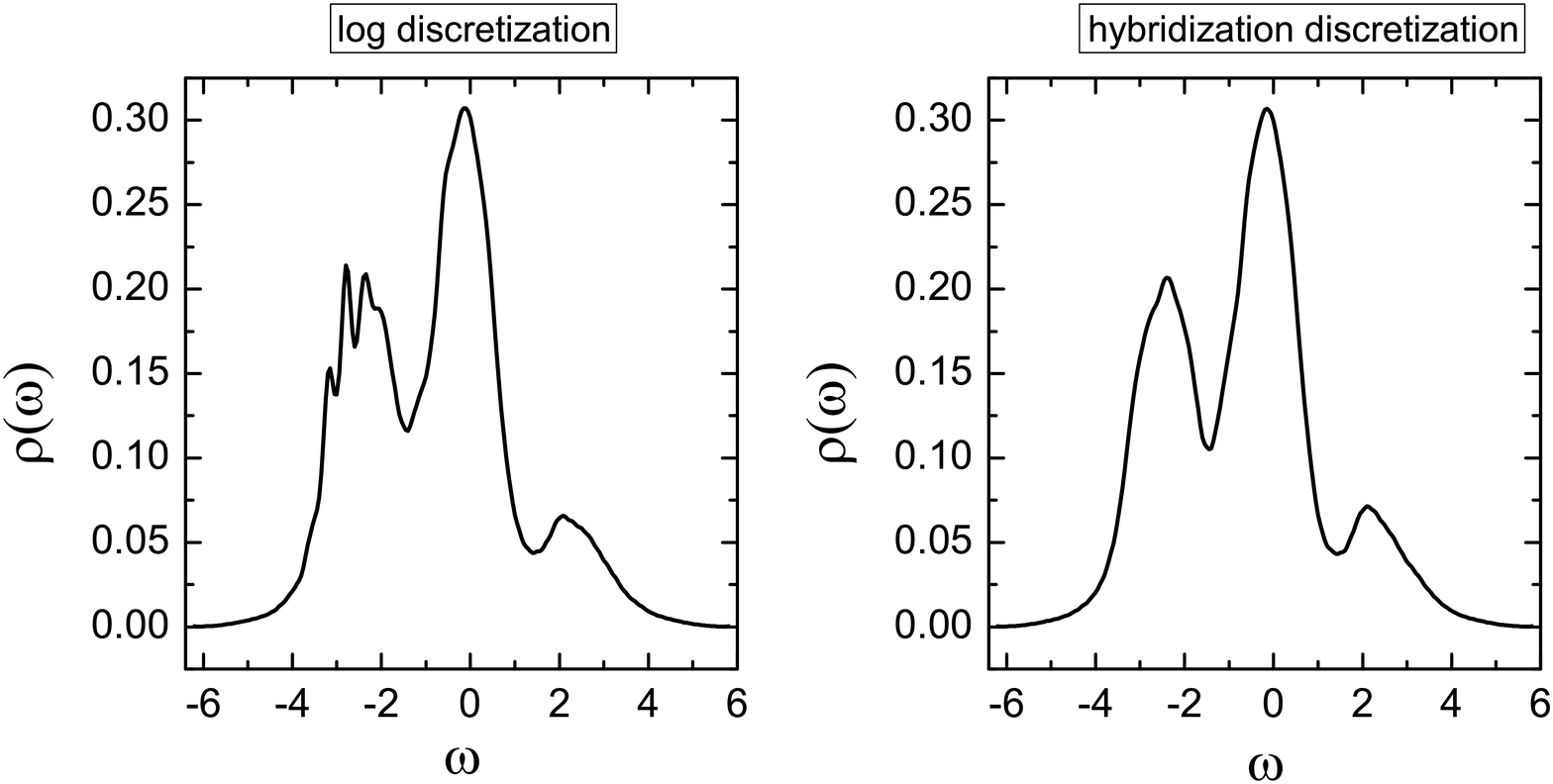}
    \caption{Comparison of different bath discretization scheme for spectral densities: (left) logarithmic discretization and (right) mixed discretization.
    For logarithmic discreitization, we choose $\Lambda=1.2$. For mixed scheme, we choose $\omega^{hyd}=0.45$ and $\Lambda=1.4$.
    We choose single impurity of Anderson impurity model with chain length $L=80$ fermionic sites, solving DMRG using kept state $M=128$.
    We select parameters as $U=3.0$ and $h=0.2$. Here we show spectral density for spin-up electrons.   }\label{sfig:compdisc}
  \end{center}
\end{figure}

\section{Deconvolution scheme}
\label{sec:decon}
In DMRG calculation, we have to introduce a broadening factor parameter $\eta$ in the retarded Green's function.
Physically, this broadening factor removes the singularity on real frequency axis. Numerically,
this broadening factor makes each single energy level is broadened by a Lorentzian peak with half-width $\eta$.
To reach the intrinsic physics, we need a scheme to extract the behavior at purely real frequencies or $\eta\rightarrow 0$.
That is, the Green's function on real frequency axis takes
\begin{equation*}
  G^R(\omega)=\lim_{\eta \rightarrow 0^{+}} G(\omega+i\eta),
\end{equation*}
where $G(\omega+i\eta)$
is calculated by dynamical DMRG introduced in Sec. \ref{sec:dmrg}.
Thus the intrinsic spectral density is
\begin{equation*}
  \rho(\omega)=-\frac{1}{\pi}\Im G_R(\omega)= \lim_{\eta \rightarrow 0^{+}} \Im G(\omega+i\eta)
\end{equation*}

Here, we use a generalized scheme, maximal entropy method,~\cite{Raas04,Raas05} to extract the information on the spectral density $\rho(\omega)$.
Let us assume that, the spectral density $g_i$ from DMRG at given values of $\omega = \xi_i$ for finite values of $\eta$ has the relation with intrinsic spectral density $\rho(\omega)$:
\begin{equation}\label{decon:raw}
  g_i= -\frac{1}{\pi}\Im G(\xi_i+i\eta) =\int L_\eta (\xi_i-\omega)\rho(\omega) =  \frac{1}{\pi} \int  d\omega \frac{\eta}{(\xi_i-\omega)^2+\eta^2} \rho(\omega)
\end{equation}
Hence the necessary step for retrieving $\rho(\omega)$ is usually called deconvolution process.


Maximal entropy method is to obtain  a continuous, non-negative spectral
density $\rho(\omega)$, which is consistent with the numerically determined values of the raw data
$\{g_i\}$. The advantage of maximal entropy method is completely unbiased. That means it does not use
any information other than the one provided by the raw data. The information content of
a density $\rho(\omega)$ is measured up to a constant by its negative entropy
\begin{equation}
  -S=\int_{-\infty}^{\infty}d\omega \rho(\omega) \ln \rho(\omega) \;.
\end{equation}
The least biased ansatz is the one with the least information content which is still compatible
with the raw data. Hence we have to look for the density $\rho(\omega)$ which minimizes
$-S$ (maximizes S) under the conditions Eq.~(\ref{decon:raw}) given by the raw data $\{g_i\}$.
To find this least biased ansatz is a straightforward task. Using the Lagrange
multipliers $\lambda_i$ for the $p$ conditions set by the raw data $\{g_i\}$
the least biased ansatz is characterized by $\delta S = 0$
\begin{equation*}
\text{min}\{-S+ \sum_i\lambda_i(\int L_\eta (\xi_i-\omega)\rho(\omega)-g_i)\} \Rightarrow 0=-1-\ln \rho(\omega) +\sum_{i=1}^{p} \lambda_i
L_{\eta}(\omega-\xi_i) \;. 
\end{equation*}

This equation implies that the least biased ansatz reads
\begin{equation}\label{decon:res}
  \rho(\omega)=\exp[-1 + \sum_{i=1}^{p} \lambda_i L_{\eta}(\omega-\xi_i) ] \;.
\end{equation}
The Lagrange multipliers are determined by the non-linear
equations Eq.~(\ref{decon:raw}):
\begin{equation}\label{}
  g_j=\sum_{n=1}^{N_{mesh}} \frac{\eta/\pi}{(\xi_j-\omega_n)^2+\eta^2}\exp[-1+\sum_{i=1}^{p}\lambda_i\frac{\eta/\pi}{(\xi_i-\omega_n)^2+\eta^2}]\;.
\end{equation}
They can be solved by non-linear equations solver package (for example, MINIPACK \cite{minipack}). Via the ansatz Eq.~(\ref{decon:res}),
the $p$ Lagrange multipliers determine
the most unbiased spectral density $\rho(\omega)$ which is still compatible with the numerically
measured information on $\rho(\omega)$.

\section{Finite-size analysis}\label{sec:size}
In this paper, the main results are based on the Anderson impurity model enclosing $L=80$ fermion sites. One natural question is
whether or not the physical results depends on system sizes. Here we briefly compare the spectral densities obtained from different system sizes.
As shown in Fig. \ref{sfig:size}, the obtained spectral densities from system size for $L=80$, $L=120$ and $L=160$ completely merge together,
which shows the system size does not influence the calculations here.
We also partly checked other parameters and confirmed that, there is no significant difference between solution from $L=80$ to $L=160$.
Thus, we conclude that a system size of $L=80$ sites is already sufficient to quantitatively capture the features of the spectral properties for the current problem.

\begin{figure}
    \includegraphics[width=0.7\columnwidth,angle=0]{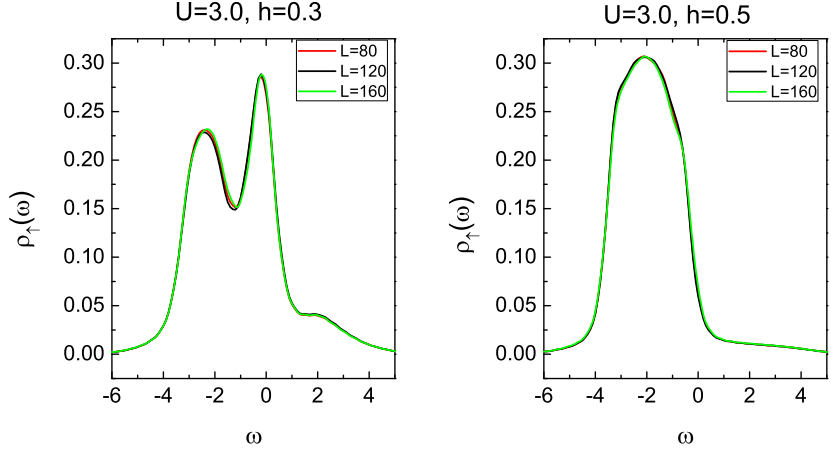}
    \caption{Spectral densities of single-orbital Hubbard model on the Bethe lattice obtained by DMFT+DMRG scheme.
    We choose the impurity model enclosing $L=80$ fermionic sites (black), $L=120$ fermionic sites (red) and $L=160$ fermionic sites (green).}\label{sfig:size}
\end{figure}

\section{Computational performance of two-component spinless model}
\label{sec:time}

\begin{figure}[!htb]
  \begin{center}
    \includegraphics[width=0.4\columnwidth,angle=0]{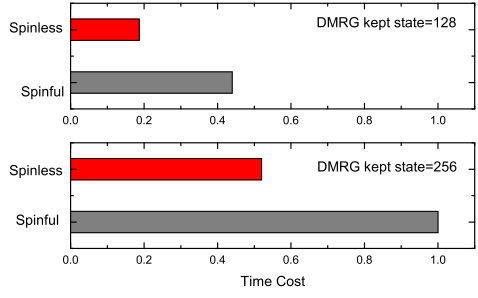}
    \caption{ CPU times for a DMRG run performed with spinful Hubbard model and two-component spinless model.
    We set all parameters in single impurity Anderson impurity model are the same.
    The time unit is the time cost of each DMFT loop for spinful Hubbard model by solving DMRG using kept state $M=256$.    }\label{sfig:cputime}
  \end{center}
\end{figure}

In this Appendix, we compare the computational performance
of the spinful Hubbard model and two-component spinless model
using DMRG calculation. We set all of physical and simulation
parameters the same for two different models.
The convergence of DMRG calculations is shown for the finite size $L=40$ for spinful Hubbard model
(equivalent two-component spinless modle enclosing $80$ fermion sites)
by keeping the number of states $M=128$ or $256$ kept in the reduced basis set, respectively.
Figure~\ref{sfig:cputime}
shows the comparison between the two models for each DMFT loop.
It is found that the two-component spinless model is faster than the spinful model by an approximate factor of $2.0$.
Furthermore, we point out that the two-component spinless model is not only faster,
but also provides a better converged resolution. In our extensive test, DMRG kept state $M=128$
is sufficient for two-component spinless model to reach converged results for most of cases. However, for the intermediate interaction regime,
the spinful Hubbard model has not even converged using DMRG kept state up to $M=256$, despite having used a
large amount of CPU time and iterations.
The key reason is that, adding a single spinless site instead of a spinful site in each DMRG step leads to a much smaller truncation error or higher resolution
(see Sec. \ref{sec:mapping} for discussion).

\section{Self-energy function}
In the main text, we have shown the spectral density which directly relates to the full local Green function $G_\sigma(\omega)$.
Here we show the related self-energy function $\Sigma(\omega)$. Generally, the self-energy can be obtained by
the Dyson equation through $\Sigma(\omega)=G^{-1}_0(\omega)-G^{-1}(\omega)$. Here we choose an alternative way to calculate
self-energy function, which was first proposed by Bulla.~\cite{bulla98}  This method of calculating self-energy turns
out to be considerably more reliable and accurate than via Dyson equation alone.

Following Bulla,~\cite{bulla98} we can calculate the quantity $F(\omega)$ first:
\begin{eqnarray}
  F_\sigma(\omega)= \langle 0 | \hat d_{\sigma}(n_{-\sigma}-1/2) \frac{1}{\omega+i\eta +E_0 -\hat H}   \hat d^{\dagger}_\sigma |0 \rangle +
  \langle 0 | \hat d^{\dagger}_{\sigma} \frac{1}{\omega+i\eta -E_0 +\hat H}   \hat d_\sigma(n_{-\sigma}-1/2) |0 \rangle.
\end{eqnarray}
and then the self-energy function is obtained by
\begin{eqnarray}
  \Sigma_\sigma(\omega)= U \frac{F_\sigma(\omega)}{G_\sigma(\omega)}.
\end{eqnarray}

We show the obtained self-energy function of spin-up electrons at impurity site in Fig.~\ref{sfig:selfenergy}.
Since the magnetic field breaks particle-hole symmetry,
the self-energy is not symmetric. And the imaginary part of self-energy shows a asymmetric two-peak structure. With increasing
magnetic field, the peak in the hole regime becomes more visible and the other peak in the electron regime tends to diminish.
That means, after phase transition, interaction only modifies the filled band electrons.

While imaginary part of self-energy relates to quasiparticle lifetime, the real part of self-energy
reflects the quasiparticle weight or effective mass.
Here we define the quasiparticle weight $Z_\sigma$,
which describe how good the single particle picture works:
\begin{equation}
Z_\sigma=[1-\frac{\partial \Re\Sigma_\sigma(\omega)}{ \partial \omega} |_{\omega=0}]^{-1}.
\end{equation}
Interestingly, the inverse of the quasiparticle weight $Z^{-1}_\sigma$ corresponds to the enhancement of the effective mass $m^{e}_\sigma(h)$ by
\begin{equation}
m^e_\sigma/m^e_0=Z^{-1}_\sigma.
\end{equation}

\begin{figure}[!htb]
    \includegraphics[width=0.7\columnwidth,angle=0]{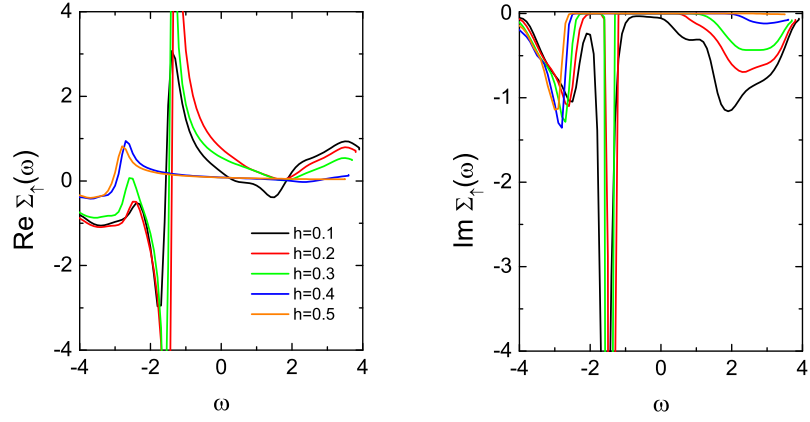}
    \caption{Self-energy function of single-orbital Hubbard model on the Bethe lattice obtained by DMFT scheme.
    Interaction strength is set to be $U=3.0$.
    We choose the impurity model enclosing $L=80$ fermionic sites, which is solved by DMRG algorithm by limiting each DMRG block with dimension $M=128$.}\label{sfig:selfenergy}
\end{figure}

The inverse of quasiparticle weight, shown in Fig. \ref{sfig:weight}, shows a sharp rise before entering the insulator phase.
This signals the enhancement of quasiparticle weight around the quantum phase transition.
This behavior provides another evidence of metamagnetic phase transition in this system.
When the ground state is fully polarized, quasiparticle weight should approach $Z^{-1}_\sigma\approx 1.0$,
corresponding to the band insulator discussed in the main text.

\begin{figure}[!htb]
    \includegraphics[width=0.45\columnwidth,angle=0]{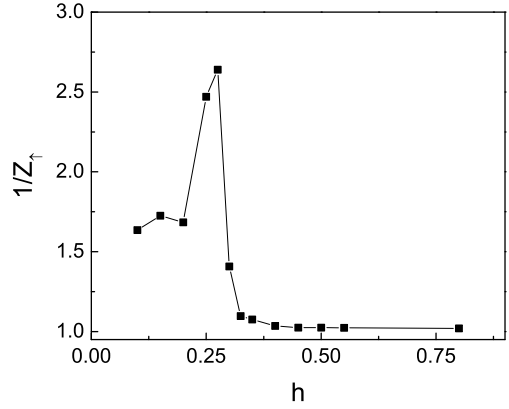}
    \caption{Inverse of quasi-particle weight as a function of magnetic field $h$, by setting $U=3.0$.}\label{sfig:weight}
\end{figure}

\end{appendices}

\end{widetext}


\begin{thebibliography}{99}
\bibitem{note1} In Ref. \onlinecite{Raas04}, the authors proposed a similar scheme by mapping the spinful fermionic model to spin model by Wigner-Jordan transformation,
which is different from the two-component mapping in this work.
\end{thebibliography}
\end{document}